\documentclass[aps,prd,10pt,showpacs,amsmath,showkeys,twocolumn,floatfix,amssymb, preprintnumbers, nofootinbib, superscriptaddress]{revtex4}
\usepackage{graphicx}
\usepackage{comment}
\usepackage[usenames]{color}
\usepackage{bm}
\usepackage{ifpdf}
\usepackage{floatrow}
\usepackage{makecell}
\usepackage{caption}

\usepackage[normalem]{ulem}
\usepackage[dvipsnames]{xcolor}
\usepackage[utf8]{inputenc}
\usepackage{hyperref}
\hypersetup{
  pdfnewwindow=true,      % links in new window
  colorlinks=true,        % false: boxed links; true: colored links
  linkcolor=PineGreen,    % color of internal links
  citecolor=PineGreen,    % color of links to bibliography
  filecolor=PineGreen,    % color of file links
  urlcolor=PineGreen      % color of external links
}
\newcommand{\be}{\begin{eqnarray}}
\newcommand{\ee}{\end{eqnarray}}

\begin{document}

\title{Charged particles interaction in both a finite volume and a uniform magnetic field II: topological and analytic properties of a magnetic system}

\author{Peng~Guo}
\email{pguo@csub.edu}

\affiliation{Department of Physics and Engineering,  California State University, Bakersfield, CA 93311, USA}

\author{Vladimir~Gasparian}
\affiliation{Department of Physics and Engineering,  California State University, Bakersfield, CA 93311, USA}

\date{\today}

\begin{abstract}
In present work, we discuss some topological features of charged particles  interacting a uniform magnetic field in a finite volume. The edge state solutions are presented, as a signature of non-trivial topological systems, the energy spectrum of edge states show up in the gap between allowed energy bands.  By treating total momentum of two-body system   as a continuous distributed parameter in complex plane, the analytic properties of solutions of finite volume system in a magnetic field is also discussed.
\end{abstract}

\maketitle

\section{Introduction}\label{sec:intro}

Study of few-body hadron/nuclear particles interaction and properties of few-body resonances  is one of important subjects in modern physics.   Hadron/nuclear particles provide the only means of understanding quantum Chromodynamics (QCD), the underlying theory of quark and gluon interactions. However, making prediction of hadron/nuclear particle interactions from first principles is not always straightforward, due to the fact that most of theoretical computations are performed in various traps, for instance,   periodic cubic box in  lattice quantum Chromodynamics (LQCD).  As the result of trapped systems, only discrete energy spectrum is observed instead of scattering amplitudes.     Hence, extracting infinite volume scattering amplitudes from discrete energy spectrum    in a trapped system  have became an important subject in   LQCD and nuclear physics communities,  see e.g. Refs.~\cite{Luscher:1990ux,Rummukainen:1995vs,Christ:2005gi,Bernard:2008ax,He:2005ey,Lage:2009zv,Doring:2011vk,Guo:2012hv,Guo:2013vsa,Kreuzer:2008bi,Polejaeva:2012ut,Hansen:2014eka,Mai:2017bge,Mai:2018djl,Doring:2018xxx,Guo:2016fgl,Guo:2017ism,Guo:2017crd,Guo:2018xbv,Mai:2019fba,Guo:2018ibd,Guo:2019hih,Guo:2019ogp,Guo:2020wbl,Guo:2020kph,Guo:2020iep,Guo:2020ikh,Guo:2020spn,Guo:2021lhz,Guo:2021uig,Guo:2021qfu,Busch98,Stetcu:2007ms, Stetcu:2010xq, Rotureau:2010uz,Rotureau:2011vf,Luu:2010hw,Yang:2016brl,Johnson:2019sps,Zhang:2019cai, Zhang:2020rhz}.  When the size of a trap is much larger than the range of interactions, the short-range particles dynamics and long-range correlation effect due to the  trap can be factorized. 
      The connection between trapped system and infinite volume system are found in a closed form, 
 \begin{equation}
   \det \left [  \cot \delta  (E) -  \mathcal{M} (E ) \right ]=0 \, ,
    \label{eqn:generalQC}
\end{equation}
   where  $\delta (E)$ refers to the diagonal matrix of infinite volume scattering phase shifts, and the   matrix function $ \mathcal{M} (E ) $  is associated to the geometry and dynamics of trap itself.  The formula that has the form of Eq.(\ref{eqn:generalQC}) is known as L\"uscher formula  \cite{Luscher:1990ux}  in LCQD  and  Busch-Englert-Rza\.zewski-Wilkens (BERW) formula \cite{Busch98} in a harmonic oscillator trap in nuclear physics community.

 In preceding work \cite{Guo:2021lhz}, a L\"uscher  formula-like formalism was presented for a finite volume two-particle system in  a uniform magnetic field.  To preserve translation symmetry and satisfy periodicity of cubic lattice, the magnetic field must be quantized and given by $\frac{2\pi}{L^2} \frac{n_p}{n_q}$, where $n_p$ and $n_q$ are integers and relatively prime  to each other, and $L$ denotes the size of cubic lattice. In LQCD, few-particle systems in a uniform magnetic field can be studied by  using  background-field methods in lattice QCD \cite{Detmold:2004qn,Detmold:2004kw,Detmold:2008xk,Detmold:2009fr}. Finite volume   formalism of a magnetic system thus may be    be used to determine the coefficient of the leading local four-nucleon operator in the process of neutral- and charged-current break-up of the deuteron. In addition to LQCD, the few-body interactions in a periodic optical lattice structure may also be simulated and studied by using ultra-cold atoms technology in atomic physics.  In condensed matter physics, it is well-known fact \cite{KOHMOTO1985343} that the magnetic field forces a wavefunction of Bloch particles to develop vortices in $\mathbf{ k}$-space, where $\mathbf{ k}$ represents the crystal momentum of a Bloch particle. The phase of wavefunction of Bloch particles is not well-defined throughout entire magnetic Brillouin zone, which is associated to a non-trivial topology of a magnetic system. When crystal momentum of a Bloch particle is forced to circle around the vortices, non-zero vorticity is ultimately related to quantized Hall conductance \cite{KOHMOTO1985343,PhysRevLett.49.405}. One of consequences of non-trivial topology of a magnetic system is the appearance of gapless topological edge states that show up in the gap between allowed energy bands \cite{PhysRevLett.71.3697,PhysRevB.48.11851,Hatsugai_1997}.     Hence  on the one hand  the system behaves as an insulator  due to the confined  circular motion of electrons in the bulk by the strong magnetic field. On the other hand,  delocalized edge states on the surface make sure that the system can conduct along the surface. The gapless edges states   are associated with the Chern number, which characterizes the topology of filled bands in two-dimensional lattice systems and sheds light on their topological properties of the system.  The potential applications of   quantum Hall insulators range from precision measurements to quantum information processing and spintronics \cite{RevModPhys.82.3045}. Ultra-cold quantum gases are a promising experimental platform to explore these effects,    the realization of topologically nontrivial band structures and artificial gauge fields \cite{Lin_2011} may be feasible. In cold atom systems, the Chern number was measured for the Hofstadter model \cite{Aidelsburger_2014}   and for the Haldane model  \cite{Asteria_2019}.

 In present work, as a continuation to our previous work  in \cite{Guo:2021lhz}, we explore  and discuss some topological and analytic properties of a finite volume two-body system in a uniform magnetic field. The total momentum of two-body system is   treated as a continuous parameter that is analogous to the crystal momentum  $\mathbf{ k}$ in condensed matter physics, the continuous distribution of a $\mathbf{ k}$ vector over the magnetic Brillouin zone forms a torus. Hence   the Berry phase can be introduced on a torus of magnetic Brillouin zone, which has non-zero value of $2\pi$ multiplied by an integer due to non-trivial topology of magnetic systems. The analytic solutions of edge states with a hard wall boundary condition in $x$-direction are also presented and discussed in current work. At last, by further taking  $\mathbf{ k}$ into a complex plane, the analytic properties of energy spectrum is also discussed.

The paper is organized as follows. The dynamics of finite volume systems in a uniform magnetic field is briefly summarized in Sec.~\ref{review}, the dynamics of finite volume magnetic systems in a plane is  reformulated  and  discussed further in Sec.~\ref{FVdynamics2D}. The topological features of a magnetic system, solutions of topological edge states and analytic properties of finite volume solutions are discussed and presented in Sec.~\ref{secberryphase}, Sec.~\ref{edgestates} and Sec.~\ref{analyticproperty} respectively.  The discussions and summary are given in Sec.~\ref{summary}.

\section{Finite volume Lippmann-Schwinger equation in a uniform magnetic field}\label{review} 
In this section, we only briefly summerize the dynamics of two-particle system in a uniform magnetic field, the details can be found in our previous work in  \cite{Guo:2021lhz}.

The dynamics of relative motion of two-particle system in a uniform magnetic field is described by 
the   finite volume    Lippmann-Schwinger (LS) equation,
\begin{equation}
\psi^{(L,  \frac{\mathbf{ P}_B}{2})}_{  \varepsilon}(\mathbf{ r})  = \int_{L_B^3} d \mathbf{ r}' G^{(L, \frac{\mathbf{ P}_B}{2})}_B(\mathbf{ r},\mathbf{ r}'; \varepsilon) V^{(L)}(\mathbf{ r}')\psi^{(L, \frac{ \mathbf{ P}_B}{2})}_{  \varepsilon}(\mathbf{ r}')  , \label{homogenousLSB}
\end{equation}
 where  the volume integration over the enlarged magnetic unit cell, $$n_q L\mathbf{  e}_x \times L \mathbf{ e}_y\times L \mathbf{ e}_z,$$ is defined by
 \begin{equation}
  \int_{L_B^3} d \mathbf{ r}'  = \int_{- \frac{n_q L}{2}}^{\frac{n_q L}{2}}  d r_x'   \int_{- \frac{  L}{2}}^{\frac{  L}{2}}  d r'_y  \int_{- \frac{  L}{2}}^{\frac{  L}{2}} d r'_z.
  \end{equation}
   The wavefunction satisfies the magnetic periodic boundary condition,
 \begin{equation}
 \psi^{(L, \frac{\mathbf{ P}_B}{2})}_\varepsilon(\mathbf{ r} + \mathbf{ n}_B L ) = e^{ i \frac{\mathbf{ P}_B}{2} \cdot \mathbf{ n}_B L}  e^{-i q B   r_y  \mathbf{ e}_x \cdot \mathbf{ n}_B L }   \psi^{(L, \frac{\mathbf{ P}_B}{2})}_\varepsilon(\mathbf{ r} ),\label{magneticperiodic}
 \end{equation}
 where
 \begin{equation}
 \mathbf{ n}_B = n_x n_q  \mathbf{e }_x + n_y \mathbf{ e}_y + n_z \mathbf{ e}_z, \ \ \ \  n_{x,y,z} \in \mathbb{Z},
 \end{equation}
 and
  \begin{equation}
 \mathbf{ P}_B =  \frac{2\pi}{L} \left ( \frac{n_x}{ n_q}  \mathbf{e }_x + n_y \mathbf{ e}_y + n_z \mathbf{ e}_z \right ), \ \ \ \  n_{x,y,z} \in \mathbb{Z}. \label{PBvecdiscrete}
 \end{equation}
The    finite volume magnetic Green's function $G^{(L, \frac{\mathbf{ P}_B}{2})}_B$  satisfies equation,
 \begin{align}
& \left ( \varepsilon -  \hat{H}_\mathbf{ r}   \right ) G^{(L, \frac{ \mathbf{ P}_B}{2})}_B(\mathbf{ r},\mathbf{ r}'; \varepsilon)  \nonumber \\
&= \sum_{\mathbf{ n}_B}  e^{  -i    \frac{ \mathbf{ P}_B  }{2}    \cdot  \mathbf{ n}_B L }  e^{  i   q B r_y \mathbf{ e}_x   \cdot  \mathbf{ n}_B L }  \delta(\mathbf{ r}-\mathbf{ r}' + \mathbf{ n}_B L),\label{magneticGBeq}
 \end{align}
  where the Hamiltonian of relative motion of two charged particles in a uniform magnetic field  is given by
  \begin{equation}
  \hat{H}_\mathbf{ r} = -   \frac{\left (\nabla_\mathbf{ r}+ i q \mathbf{ A} (\mathbf{ r}) \right )^2}{2\mu}  .
 \end{equation}
  $ q  $ and $\mu$ stand for effective charge and mass of two particles respectively.   The uniform magnetic field is chosen   along $z$-axis, $\mathbf{ B}= B \mathbf{ e}_z$, and Landau gauge for vector potential is adopted in this work,
\begin{equation}
\mathbf{ A} (\mathbf{ x} ) = B (0, x, 0).
\end{equation} 
To warrant a state that is translated through a closed path remain same, the magnetic flux $qB n_q L^2$ through the surface of an enlarged magnetic  unit cell in $x-y$ plane must be quantized:
 \begin{equation}
 q B n_q L^2 = 2\pi  n_p , 
 \end{equation}
 where $n_p$ and $ n_q$ are two relatively prime integers.

The analytic   solutions of $G^{(L, \frac{\mathbf{ P}_B}{2})}_B$  can be constructed from its infinite volume counterpart $G^{(\infty)}_B$ by, 
\begin{align}
& G^{(L,  \frac{\mathbf{ P}_B}{2})}_B(\mathbf{ r},\mathbf{ r}'; \varepsilon) \nonumber \\
& = \sum_{\mathbf{ n}_B}    G^{(\infty)}_B(\mathbf{ r},\mathbf{ r}' + \mathbf{ n}_B L; \varepsilon)   e^{  i    \frac{ \mathbf{ P}_B  }{2}    \cdot  \mathbf{ n}_B L }   e^{  -i   q B r'_y \mathbf{ e}_x   \cdot  \mathbf{ n}_B L }  \nonumber \\
& = \sum_{\mathbf{ n}_B}     e^{ - i    \frac{ \mathbf{ P}_B  }{2}    \cdot  \mathbf{ n}_B L }   e^{  i   q B r_y \mathbf{ e}_x   \cdot  \mathbf{ n}_B L } G^{(\infty)}_B(\mathbf{ r} + \mathbf{ n}_B L,\mathbf{ r}' ; \varepsilon)  , \label{magneticGreendef}
\end{align}
where the infinite volume magnetic Green's function $G^{(\infty)}_B$ satisfies equation,
  \begin{equation}
 \left ( \varepsilon -  \hat{H}_\mathbf{ r}   \right ) G^{(\infty)}_B(\mathbf{ r},\mathbf{ r}'; \varepsilon)  =  \delta(\mathbf{ r}-\mathbf{ r}'  ),
 \end{equation}
The 3D analytic expression of $G^{(\infty)}_B$ is related to 2D magnetic Green's function that is defined in $x-y$ plane, $G^{(\infty,2D)}_B$, by
 \begin{align}
 & G^{(\infty)}_B(\mathbf{ r},\mathbf{ r}'; \varepsilon)    =  \int_{- \infty}^\infty \frac{d p_z}{2\pi}G^{(\infty,2D)}_B( \bm{\rho}  , \bm{\rho}'; \varepsilon- \frac{p_z^2}{2\mu })      e^{i p_z (r_z - r'_z)}  , \label{G3DtoG2D}
 \end{align}
 where $$\bm{\rho} = r_x \mathbf{ e}_x + r_y \mathbf{ e}_y, \ \ \ \ \bm{\rho}' = r'_x \mathbf{ e}_x + r'_y \mathbf{ e}_y$$ are relative coordinates defined in $x-y$ plane.  The various representations of 2D infinite volume magnetic Green's function, $G^{(\infty,2D)}_B$, are  given  by
 \begin{align}
 & G^{(\infty,2D)}_B(\bm{\rho}  , \bm{\rho}';  \varepsilon)  \nonumber \\
 &=\sum_{n=0}^\infty \int_{- \infty}^\infty \frac{d p_y}{2\pi}   \frac{ \phi_n(r_x + \frac{p_y}{qB}) \phi^*_n(r'_x + \frac{p_y}{qB}) e^{i p_y (r_y - r'_y)} }{\varepsilon - \frac{qB}{\mu}(n+ \frac{1}{2})   } \nonumber \\
 &   =  \frac{ qB}{2\pi}  e^{ - \frac{i q B}{2} (r_x + r'_x) (r_y - r'_y)}       \sum_{n=0}^\infty \frac{ L_n ( \frac{qB}{2}  | \bm{\rho} - \bm{\rho}' |^2 )   e^{- \frac{qB}{4}  | \bm{\rho} - \bm{\rho}' |^2  }  }{  \varepsilon - \frac{qB}{\mu} (n+ \frac{1}{2})  }    \nonumber \\
&  =-   \frac{ 2 \mu }{4 \pi}   e^{ - \frac{i q B}{2} (r_x + r'_x) (r_y - r'_y)} e^{- \frac{qB}{4}  | \bm{\rho} - \bm{\rho}' |^2  }  \Gamma(\frac{1}{2} - \frac{\mu \varepsilon}{q B})  \nonumber \\
& \quad   \times U (\frac{1}{2} - \frac{\mu \varepsilon}{q B}, 1,  \frac{qB}{2}  | \bm{\rho} - \bm{\rho}' |^2  ), \label{G2Dinf}
 \end{align}
where  $  \phi_n(r_x) $ is eigen-solution of 1D harmonic oscillator potential,
 \begin{equation}
  \phi_n(r_x)  = \frac{1}{\sqrt{2^n n!}} \left ( \frac{qB}{\pi} \right )^{\frac{1}{4}} e^{- \frac{q B}{2} r_x^2} H_n (\sqrt{qB} r_x).
 \end{equation}
$H_n(x)$, $L_n (x)$ and $U(a,b,z)$ are standard Hermite polynomial, Laguerre polynomial and Kummer function respectively \cite{NIST:DLMF} .

\section{Finite volume dynamics of a magnetic system in a plane}\label{FVdynamics2D} 

From this point on,  all the discussions will be   restricted   in $x-y$ plane, the purpose of this is only to simplify the  technical  presentations. The conclusions can in principle be extended into 3D as well by using relation in Eq.(\ref{G3DtoG2D}). In this section, the dynamical equations of a magnetic system in a plane will be reformulated in terms of new basis functions that satisfy magnetic periodic boundary conditions. In terms of these magnetic periodic basis functions,   reaction amplitudes may be introduced, and LS equation of reaction amplitudes is thus obtained.  The relation to Harper's equation is presented when a specific type of potential is considered. At last,  the quantization conditions in 2D plane with contact interactions are presented and discussed.

In $x-y$ plane, similar to Eq.(\ref{homogenousLSB}), the   finite volume    LS equation  in 2D  is given by
\begin{align}
&\psi^{(L, \frac{ \mathbf{ P}_B}{2},2D)}_{  \varepsilon}( \bm{\rho})   \nonumber \\
&= \int_{L_B^2} d  \bm{\rho}' G^{(L,  \frac{\mathbf{ P}_B}{2},2D)}_B(\bm{\rho},\bm{\rho}'; \varepsilon) V^{(L)}(\bm{\rho}')\psi^{(L, \frac{ \mathbf{ P}_B}{2},2D)}_{  \varepsilon}(\bm{\rho}')  , \label{homogenousLSB2D}
\end{align}
 where  the volume integration over the magnetic unit cell is defined by
 \begin{equation}
  \int_{L_B^2} d \bm{ \rho}'  = \int_{- \frac{n_q L}{2}}^{\frac{n_q L}{2}}  d r_x'   \int_{- \frac{  L}{2}}^{\frac{  L}{2}}  d r'_y   .
  \end{equation}
  One of analytic expression of finite volume 2D magnetic Green's function is explicitly given by
  \begin{align}
 & G^{(L, \frac{\mathbf{ P}_B}{2},2D)}_B(\bm{\rho}  , \bm{\rho}';  \varepsilon)  \nonumber \\
 & =  \sum_{n=0}^\infty \sum_{n_x \in \mathbb{Z}} e^{- i (\frac{P_{Bx}}{2} - q B r_y) n_x n_q L} \frac{1}{L} \sum_{\substack{ p_y = \frac{2\pi n_y}{L}+ \frac{P_{By}}{2},  n_y \in \mathbb{Z} } } \nonumber \\
 &\times   \frac{ \phi_n(r_x + n_x n_q L + \frac{p_y}{qB}) \phi^*_n(r'_x + \frac{p_y}{qB}) e^{i p_y (r_y - r'_y)} }{\varepsilon - \frac{qB}{\mu}(n+ \frac{1}{2})   }  . \label{GLB2Dphin}
 \end{align}
By splitting lattice sum of $n_y$ in $k_y$ in Eq.(\ref{GLB2Dphin})  using identity,
\begin{equation}
\sum_{  n_y \in \mathbb{Z}  }  f(n_y)= \sum_{\alpha=0}^{n_p-1} \sum_{    \overline{n}_y  \in \mathbb{Z}  }  f(n_p \overline{n}_y + \alpha ),
\end{equation}
and also  performing a shifting in lattice sum of $n_x$: $n_x  \rightarrow n_x - \overline{n}_y$,  the finite volume 2D magnetic Green's function thus can also be    written as
   \begin{equation}
  G^{(L, \frac{ \mathbf{ P}_B }{2},2D)}_B(\bm{\rho}  , \bm{\rho}';  \varepsilon)  = \sum_{n=0}^\infty  \sum_{\alpha=0}^{n_p-1}     \frac{     \chi_{n,\alpha}^{( \frac{ \mathbf{ P}_B}{2})}(\bm{\rho} ) \chi_{n,\alpha}^{(\frac{\mathbf{ P}_B}{2}) *}(\bm{\rho}' ) }{\varepsilon - \frac{qB}{\mu}(n+ \frac{1}{2})   }  , \label{GB2Dchin}
 \end{equation}
 where
  \begin{align}
 & \chi_{n,\alpha}^{(\frac{\mathbf{ P}_B}{2})}(\bm{\rho} )  \nonumber \\
 &= \frac{1}{\sqrt{L}}\sum_{n_x \in \mathbb{Z}}  \phi_n(r_x  + \frac{ \frac{2\pi  (n_p  n_x + \alpha)}{L} + \frac{P_{By}}{2} }{qB})  e^{- i \frac{P_{Bx}}{2}   n_x n_q L} \nonumber \\
 & \times   e^{i (\frac{2\pi  (n_p  n_x+ \alpha) }{L}+ \frac{P_{By}}{2}) r_y  } . \label{chidef}
 \end{align}
The $ \chi_{n,\alpha}^{(\frac{\mathbf{ P}_B}{2})}(\bm{\rho} ) $ functions are   solutions of Schr\"odinger equation with degeneracy of $n_p$ for a fixed $n$ value, 
   \begin{align}
 \left (  \frac{qB}{\mu} (n+ \frac{1}{2}) -  \hat{H}_{\bm{ \rho}}   \right ) &  \chi_{n,\alpha}^{(\frac{\mathbf{ P}_B}{2})} (\bm{ \rho} ) =0,  \nonumber \\
 \alpha & = 0, 1, \cdots, n_p -1,
 \end{align}
  and they too satisfy magnetic periodic boundary condition,
 \begin{equation}
\chi_{n,\alpha}^{(\frac{\mathbf{ P}_B}{2})} (\bm{ \rho} + \mathbf{ n}_B L ) = e^{ i \frac{\mathbf{ P}_B}{2} \cdot \mathbf{ n}_B L}  e^{-i q B   r_y  \mathbf{ e}_x \cdot \mathbf{ n}_B L }   \chi_{n,\alpha}^{(\frac{\mathbf{ P}_B}{2})} (\bm{ \rho} ).\label{chiperiodicbd}
 \end{equation}

 Using orthogonality relation of 1D harmonic oscillator basis functions,
 \begin{equation}
\int_{-\infty}^\infty d r_x \phi_n(r_x) \phi^*_{n'} (r_x) = \delta_{n, n'},
\end{equation}
and also completeness of 1D harmonic oscillator basis
\begin{equation}
 \sum_{n=0}^\infty  \phi_n(r_x) \phi^*_{n} (r'_x)  =  \delta(r_x - r'_x),
\end{equation}
one can show easily that $ \chi_{n,\alpha}^{(\mathbf{ P}_B)}(\bm{\rho} ) $ functions are  orthogonal,
 \begin{equation}
    \int_{L_B^2} d \bm{ \rho}    \chi_{n,\alpha}^{(\frac{\mathbf{ P}_B}{2})}(\bm{\rho} ) \chi_{n',\alpha'}^{(\frac{\mathbf{ P}_B}{2}) *}(\bm{\rho} )  =  \delta_{\alpha, \alpha'}   \delta_{n, n'} , \label{chiorthogonal}
\end{equation}
and   form a complete magnetic periodic basis as well,
\begin{align}
& \sum_{n=0}^\infty  \sum_{\alpha =0}^{n_p-1}   \chi_{n,\alpha}^{(\frac{\mathbf{ P}_B}{2})}(\bm{\rho} ) \chi_{n,\alpha}^{( \frac{\mathbf{ P}_B}{2}) *}(\bm{\rho}' )  \nonumber \\
& =   \sum_{\mathbf{ n}_B  }  e^{- i \frac{\mathbf{ P}_{B}}{2} \cdot    \mathbf{ n}_B L} e^{i  q B   r_y \mathbf{ e}_x \cdot \mathbf{ n}_B L   }      \delta(\bm{\rho} - \bm{\rho}' +  \mathbf{ n}_B L  )       .
\end{align}
 Therefore, in presence of magnetic field, it is more convenient to use $ \chi_{n,\alpha}^{(\frac{\mathbf{ P}_B}{2})}(\bm{\rho} ) $ as basis functions  instead of   plane wave, such as $e^{i \mathbf{ k} \cdot \bm{\rho}}$ with $\mathbf{ k} = \frac{2\pi \mathbf{ n}}{L} + \frac{\mathbf{ P}}{2}$ and $\mathbf{ n} \in \mathbb{Z}^2$, which are common choice in finite volume.

\subsection{Finite volume reaction amplitudes of a magnetic system}

In absence of magnetic field,  the momentum representation of finite volume LS equation normally has the form of
\begin{equation}
T^{(\frac{\mathbf{ P}}{2})}_{\mathbf{ p}}(\varepsilon) = \frac{1}{L^2} \sum_{\mathbf{ p}'  } \frac{\widetilde{V}_{\mathbf{ p},\mathbf{ p}'}}{\varepsilon - \frac{{\mathbf{ p}'}^2}{2\mu}} T^{(\frac{\mathbf{ P}}{2})}_{\mathbf{ p}'}(\varepsilon) , 
\end{equation}
where $$(\mathbf{ p} ,\mathbf{ p}') \in   \frac{2\pi \mathbf{ n}}{L}+ \frac{\mathbf{ P}}{2}, \ \  \mathbf{ n} \in \mathbb{Z}^2,$$  and $\mathbf{ P} = \frac{2\pi \mathbf{ d}}{L}, \mathbf{ d} \in \mathbb{Z}^2$ represents the total momentum of particles system.  The finite volume scattering amplitude $T^{(\frac{\mathbf{ P}}{2})}_{\mathbf{ p}}(\varepsilon) $ amplitudes and matrix element of potential are defined in terms of plane wave basis by
\begin{align}
T^{(\frac{\mathbf{ P}}{2})}_{\mathbf{ p}}(\varepsilon)  &= - \int_{L^2} d  \bm{\rho}'  e^{- i \mathbf{ p} \cdot \bm{\rho}}  V^{(L)}(\bm{\rho}')\psi^{(L, \frac{\mathbf{ P}}{2},2D)}_{  \varepsilon}(\bm{\rho}'),  \nonumber \\
\widetilde{V}_{\mathbf{ p},\mathbf{ p}'} & = \int_{L^2} d  \bm{\rho}  e^{- i \mathbf{ p} \cdot \bm{\rho}}   V^{(L)}(\bm{\rho})   e^{ i \mathbf{ p}' \cdot \bm{\rho}}  ,
\end{align}
see e.g. Ref.~\cite{Guo:2020kph}.

Similarly, with the magnetic field on,  the finite volume reaction amplitude may be introduced  in terms of   $ \chi_{n,\alpha}^{(\frac{\mathbf{ P}_B}{2})}(\bm{\rho} ) $ basis  functions by
\begin{equation}
T^{(\frac{\mathbf{ P}_B}{2})}_{n,\alpha}  (\varepsilon)= - \int_{L_B^2} d  \bm{\rho}'   \chi_{n,\alpha}^{(\frac{\mathbf{ P}_B}{2})*}(\bm{\rho}' ) V^{(L)}(\bm{\rho}')\psi^{(L,\frac{ \mathbf{ P}_B}{2},2D)}_{  \varepsilon}(\bm{\rho}').
\end{equation} 
Using    Eq.(\ref{GB2Dchin}),  the representation of   LS   Eq.(\ref{homogenousLSB2D}) in terms of finite volume reaction amplitudes  is  thus obtained
 \begin{equation}
T^{(\frac{\mathbf{ P}_B}{2})}_{n,\alpha}  (\varepsilon) =   \sum_{n'=0}^\infty  \sum_{\alpha'=0}^{n_p-1}     \frac{     V^{(\frac{\mathbf{ P}_B}{2})}_{n,\alpha; n', \alpha'}     }{\varepsilon - \frac{qB}{\mu}(n'+ \frac{1}{2})   }   T^{(\frac{\mathbf{ P}_B}{2})}_{n',\alpha'}  (\varepsilon) , \label{TLS2D}
\end{equation}
 where
  \begin{equation}
V^{(\frac{\mathbf{ P}_B}{2})}_{n,\alpha; n', \alpha'}   = \int_{L_B^2} d  \bm{\rho}   \chi_{n,\alpha}^{(\frac{\mathbf{ P}_B}{2})*}(\bm{\rho} ) V^{(L)}(\bm{\rho})   \chi_{n',\alpha'}^{(\frac{\mathbf{ P}_B}{2})}(\bm{\rho} )   . \label{Vmatrixelement}
\end{equation}
Hence, the energy spectrum of a magnetic system may be determined  from homogenous equation, Eq.(\ref{TLS2D}), by
\begin{equation}
\det \left [  \delta_{n,\alpha; n', \alpha'}  -  \frac{     V^{(\frac{\mathbf{ P}_B}{2})}_{n,\alpha; n', \alpha'}     }{\varepsilon - \frac{qB}{\mu}(n'+ \frac{1}{2})   }   \right ] =0.
\end{equation}

\subsection{Relation to Harper's equation}
In this subsection, we will show how  the well-known Harper's equation \cite{Harper_1955,PhysRevLett.49.405} is obtained from Eq.(\ref{TLS2D}), when
 a specific type of potential is considered,
\begin{equation}
V^{(L)}(\bm{\rho})  = V_1 \cos \frac{2\pi r_x}{L}+  V_2 \cos \frac{2\pi r_y}{L}.
\end{equation} 
Thus, the matrix element of potential term is given by
 \begin{align}
& V^{(\frac{\mathbf{ P}_B}{2})}_{n,\alpha; n', \alpha'}  \nonumber \\
 & =     \delta_{\alpha, \alpha'}        \frac{ e^{-i  \frac{ \frac{2\pi    \alpha}{L} + \frac{P_{By}}{2} }{ \frac{L}{2\pi} qB} } +  (-1)^{n+n'}  e^{i   \frac{ \frac{2\pi    \alpha}{L} + \frac{P_{By}}{2} }{\frac{L}{2\pi}  qB} }}{2}V^{(1)}_{n , n' }     \nonumber \\
 &+    \frac{ \delta_{\alpha, \alpha'+1} V^{(2, -)}_{n , n' }  +  \delta_{\alpha, \alpha'-1} V^{(2,+)}_{n , n' } }{2}     ,
\end{align}
where
\begin{align}
V^{(1)}_{n , n' }  &=V_1   \int_{ - \infty}^{ \infty} d r_x    \phi^*_n(r_x   )  e^{i \frac{2\pi r_x}{L} }   \phi_{n'}(r_x   ) , \nonumber \\
V^{(2, \pm)}_{n , n' }  &= V_2  \int_{ - \infty}^{\infty} d r_x  \phi^*_n(r_x  \pm    \frac{ 1   }{ \frac{L}{2\pi} qB})   \phi_{n'}(r_x  ) .
\end{align}
Redefining reaction amplitude by
\begin{equation}
T^{(\frac{\mathbf{ P}_B}{2})}_{n,\alpha}  (\varepsilon)  =  \left ( \varepsilon - \frac{qB}{\mu}(n+ \frac{1}{2}) \right ) d^{(\frac{\mathbf{ P}_B}{2})}_{n,\alpha}  (\varepsilon)  e^{ i \frac{P_{B x}}{2}  \frac{ \frac{2\pi }{L} \alpha }{qB} },
\end{equation}
and also adopting nearest neighbour approximation:
\begin{equation}
V^{(1)}_{n , n' } \simeq \delta_{n, n'} V^{(1)}_{n  } , \ \  \ \  V^{(2, \pm)}_{n , n' } \simeq \delta_{n, n'} V^{(2)}_{n  } ,
\end{equation}
the  LS Eq.(\ref{TLS2D})   can thus be turned into   Harper's equation \cite{Harper_1955,PhysRevLett.49.405},
 \begin{align}
&  \left ( \varepsilon - \frac{qB}{\mu}(n+ \frac{1}{2}) \right ) d^{(\frac{\mathbf{ P}_B}{2})}_{n,\alpha}  (\varepsilon)    \nonumber \\
 &  =    V^{(1)}_{n  }     \cos \left ( \frac{ \frac{2\pi    \alpha}{L} + \frac{P_{By}}{2} }{ \frac{L}{2\pi} qB}  \right )             d^{(\frac{\mathbf{ P}_B}{2})}_{n,\alpha}  (\varepsilon)  \nonumber \\
 & +    V^{(2)}_{n  }   \frac{       (\varepsilon)  e^{- i  \frac{ \frac{P_{B x}}{2}   }{ \frac{L}{2\pi} qB} }  d^{(\frac{\mathbf{ P}_B}{2})}_{n,\alpha-1} +      e^{ i  \frac{ \frac{P_{B x}}{2}   }{ \frac{L}{2\pi} qB} }    d^{(\frac{\mathbf{ P}_B}{2})}_{n,\alpha+1}  (\varepsilon)  }{2}     .
\end{align}
The Harper's equation plays a crucial role in understanding topological features of a magnetic system in condensed matter physics, see e.g. \cite{PhysRevLett.49.405}.

\subsection{Contact interaction and quantization condition}

The short-range nuclear force may be modeled by contact interaction, in this work, $S$-wave dominance is assumed, so we will adopt a simple periodic contact potential,
\begin{equation}
V^{(L)}(\bm{\rho}) = \sum_{\mathbf{ n} \in \mathbb{Z}^2} V_0  \delta(\bm{\rho} + \mathbf{ n} L) .
\end{equation}
Hence Eq.(\ref{homogenousLSB}) is reduced to matrix equation,
\begin{align}
& \psi^{(L, \frac{ \mathbf{ P}_B}{2} , 2D)}_{  \varepsilon}( \eta L \mathbf{ e}_x)   \nonumber \\
&=  \sum_{\eta' =0}^{n_q-1} G^{(L, \frac{\mathbf{ P}_B}{2},2D)}_B(\eta L \mathbf{ e}_x , \eta'  L \mathbf{ e}_x; \varepsilon) V_0  \psi^{(L, \frac{\mathbf{ P}_B}{2}, 2D)}_{  \varepsilon}( \eta' L \mathbf{ e}_x)  , 
\end{align}
where $\eta = 0, \cdots, n_q-1$, and $\eta \mathbf{ e}_x$ stand for the location of $n_q$ scattering centers placed in an enlarged magnetic cell in a plane: $n_q L \mathbf{ e}_x \times L \mathbf{ e}_y$. The eigen-energy spectrum is thus determined by quantization condition,
\begin{equation}
 \det \left [ \frac{\delta_{\eta , \eta'}}{V_0}  -  G^{(L, \frac{ \mathbf{ P}_B}{2},2D)}_B(\eta L \mathbf{ e}_x , \eta' L \mathbf{ e}_x; \varepsilon)     \right ] = 0. \label{qc2DV0}
\end{equation}
Both bare strength of potential $V_0$ and the diagonal component of finite volume magnetic Green's function, $G^{(L, \frac{ \mathbf{ P}_B}{2},2D)}_B$, are ultra-violet (UV) divergent.  Ultimately, UV divergence in both $V_0$  and $G^{(L, \frac{ \mathbf{ P}_B}{2},2D)}_B$  must be regularized and cancel out explicitly, and quantization condition in Eq.(\ref{qc2DV0})  is thus free of UV divergence  and well-defined.

\subsubsection{Scattering in infinite volume with a contact interaction}
In a infinite 2D plane, the two-body scattering by a contact interaction, 
\begin{equation}
V^{(\infty)}(\bm{\rho})=V_0 \delta (\bm{\rho}), \label{V0inf}
\end{equation}
 is described by a inhomogeneous LS equation,
\begin{equation}
\psi^{(\infty,2D)}_{   \mathbf{ k}}(  \bm{\rho}  )  = e^{i \mathbf{ k} \cdot \bm{\rho}} +  V_0 G^{(\infty,2D)}_0( \bm{\rho}  ; \varepsilon)  \psi^{(\infty,2D)}_{   \mathbf{ k}}(  \mathbf{ 0}  ), \label{inhomogenousLS}
\end{equation}
where $\mathbf{ k}$ stands for incoming relative momentum of two particles, and it is related to $\varepsilon$ by $$k^2 = 2\mu \varepsilon.$$  The infinite volume Green's function is given by
\begin{equation}
G^{(\infty,2D)}_0( \bm{\rho}  ; \varepsilon)  = \int_{-\infty}^\infty \frac{d \mathbf{p }}{(2\pi)^2}  \frac{ e^{i \mathbf{ p} \cdot \bm{\rho}}}{\varepsilon - \frac{\mathbf{ p}^2}{2\mu}} = - \frac{2\mu i}{4} H_0^{(1)} (k \rho).
 \end{equation}
The Eq.(\ref{inhomogenousLS}) can be rewritten as
\begin{equation}
\psi^{(\infty,2D)}_{   \mathbf{ k}}(  \bm{\rho}  )  = e^{i \mathbf{ k} \cdot \bm{\rho}} +    i t^{(\infty)}_0 (k) H_0^{(1)} (k \rho),
\end{equation}
where 
\begin{equation}
t^{(\infty)}_0 (k)  =  - \frac{2\mu }{4} \frac{1}{ \frac{1}{V_0} - G^{(\infty,2D)}_0(  \mathbf{ 0}  ; \varepsilon)  }
\end{equation}
represents $S$-wave two-body scattering amplitude. $t^{(\infty)}_0 (k) $ is normally parameterized by a phase shift,
\begin{equation}
t^{(\infty)}_0 (k)  =  \frac{1}{\cot \delta_0 ( \varepsilon ) -i }.
\end{equation}
Hence,  $V_0$ is related to scattering phase shift in infinite volume by
\begin{equation}
\cot \delta_0 (\varepsilon) = - \frac{4}{2\mu V_0} + \frac{2}{\pi} ( \gamma_E + \frac{1}{2} \ln \frac{\mu \varepsilon  \rho^2}{2}) |_{\rho \rightarrow 0}. \label{V0phaseshift}
\end{equation}

\subsubsection{Quantization condition of a magnetic system in infinite volume}
The eigen-energy of the charge particles system in a uniform magnetic field is in fact discretized even in infinite volume. The dynamics of a trapped system by magnetic field  in infinite volume is also described by a homogeneous equation similar to Eq.(\ref{homogenousLSB}). Hence, with a contact interaction given in Eq.(\ref{V0inf}), the quantization condition of a magnetic system in infinite volume is thus given by
\begin{equation}
  \frac{1}{V_0} =   G^{( \infty,2D)}_B( \mathbf{ 0} ,  \mathbf{ 0}; \varepsilon)     , 
\end{equation}
where 
\begin{equation}
G^{( \infty,2D)}_B( \mathbf{ 0} ,  \mathbf{ 0}; \varepsilon)   =  -   \frac{ 2 \mu }{4 \pi}   \Gamma(\frac{1}{2} - \frac{\mu \varepsilon}{q B})      U (\frac{1}{2} - \frac{\mu \varepsilon}{q B}, 1,  \frac{qB}{2}  \rho^2   )|_{\rho \rightarrow 0}.
\end{equation}
Using Eq.(\ref{V0phaseshift}) and asymptotic form of Kummer function,
\begin{align}
& -  \Gamma(\frac{1}{2} - \frac{\mu \varepsilon}{q B})   U (\frac{1}{2} - \frac{\mu \varepsilon}{q B}, 1,  \frac{qB}{2}  \rho^2   )|_{\rho \rightarrow 0}  \nonumber \\
&=  2  \gamma_E + \psi(\frac{1}{2} - \frac{\mu \varepsilon}{qB}) + \ln \frac{qB \rho^2}{2 } |_{\rho \rightarrow 0} , \label{kummerUV}
\end{align}
where $$\psi(x) = \frac{d}{d x} \ln \Gamma(x)$$ is logarithmic derivative of the Gamma function, thus, after UV cancellation, we find
\begin{equation}
   \cot \delta_0 (\varepsilon)   -  \mathcal{M}^{( \infty, 2D)}_B( \mathbf{ 0} ,  \mathbf{ 0};  \varepsilon)    =0 ,
\end{equation}
where 
\begin{equation}
 \mathcal{M}^{( \infty, 2D)}_B( \mathbf{ 0} ,  \mathbf{ 0};  \varepsilon)   =  -  \frac{1}{\pi} \left [  \psi(\frac{1}{2} - \frac{\mu \varepsilon}{qB}) + \ln \frac{ q B}{\mu \varepsilon} \right ],
\end{equation}
is UV-free and well-defined function in infinite volume.

  \begin{figure}
\begin{center}
\includegraphics[width=1.0\textwidth]{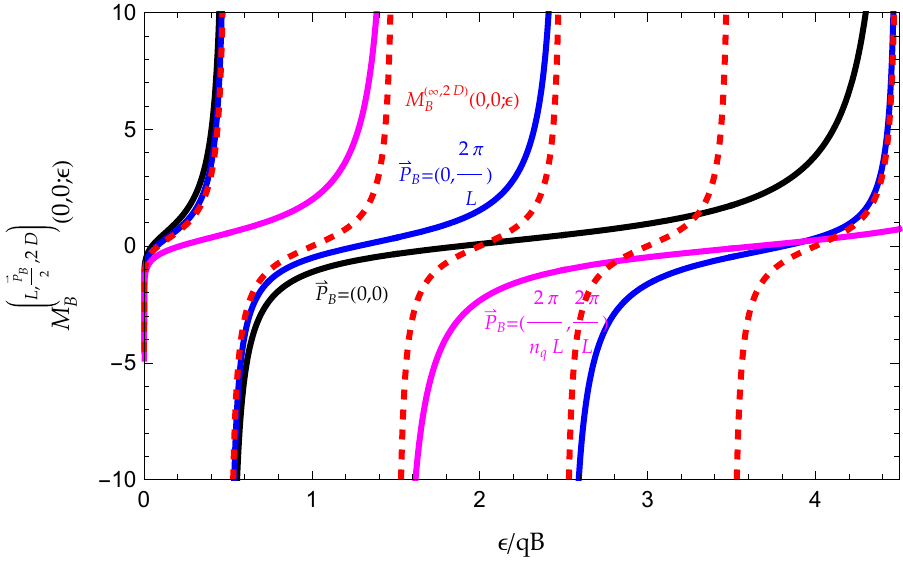}
\caption{    The plot of finite volume magnetic zeta function $ \mathcal{M}^{( L,\frac{ \mathbf{ P}_B}{2} ,2D)}_B(\mathbf{ 0} , \mathbf{ 0}; \varepsilon) $  defined in Eq.(\ref{zetafv}) vs.  $ \mathcal{M}^{(   \infty ,2D)}_B(\mathbf{ 0} , \mathbf{ 0}; \varepsilon) $ (red dahsed) given in Eq.(\ref{zetaMinf2D}).   The solid black, blue and pink curves are corresponding to $\mathbf{ P}_B = (0,0), (0, \frac{2\pi}{L}) $ and $(\frac{2\pi}{n_q L}, \frac{2\pi}{L})$ respectively.   The parameters are chosen as:      $L=5$, and $n_q=n_p=1$.}\label{Mplot}
\end{center}
\end{figure}

\subsubsection{Quantization condition in finite volume}

Using Eq.(\ref{V0phaseshift}) and also introducing the matrix elements of generalized magnetic zeta function by
\begin{align}
& \mathcal{M}^{(L,\frac{\mathbf{ P}_B}{2},2D)}_B(\eta L \mathbf{ e}_x , \eta'  L\mathbf{ e}_x; \varepsilon)  \nonumber \\
& =  - \frac{4}{2\mu} G^{(L, \frac{\mathbf{ P}_B}{2},2D)}_B(\eta L \mathbf{ e}_x , \eta' L \mathbf{ e}_x; \varepsilon)  \nonumber \\
&+ \delta_{\eta,\eta'}  \frac{1}{\pi} (\gamma_E + \frac{1}{2}\ln \frac{\mu \varepsilon \rho^2}{2}) |_{\rho \rightarrow 0}, \label{zetafv}
\end{align}
thus, the quantization condition in  Eq.(\ref{qc2DV0}) now can be recasted in a L\"uscher formula-like form \cite{Luscher:1990ux}  ,
\begin{equation}
 \det \left [  \delta_{\eta , \eta'}  \cot \delta_0 ( \varepsilon )   -  \mathcal{M}^{(L,\frac{\mathbf{ P}_B}{2},2D)}_B(\eta L \mathbf{ e}_x , \eta' L \mathbf{ e}_x; \varepsilon)    \right ] = 0, \label{qc2D}
\end{equation}
where $(\eta, \eta') = 0, \cdots , n_q -1$.

Using Eq.(\ref{magneticGreendef}) and Eq.(\ref{G2Dinf}), the generalized magnetic zeta function is thus given explicitly by
\begin{align}
& \mathcal{M}^{(L,\frac{\mathbf{ P}_B}{2},2D)}_B(\eta L \mathbf{ e}_x , \eta'  L\mathbf{ e}_x; \varepsilon)    \nonumber \\
&=   \mathcal{M}^{( \infty, 2D)}_B(\eta L \mathbf{ e}_x , \eta'  L\mathbf{ e}_x ; \varepsilon)   \nonumber \\
& +  \frac{ 1 }{\pi}     \sum_{\mathbf{ n}_B \neq \mathbf{ 0}}     e^{ - i    \frac{ \mathbf{ P}_B  }{2}    \cdot  \mathbf{ n}_B L }   e^{  i   q B  \eta n_x n_q L^2    }  \nonumber \\
& \times   e^{ - \frac{i q B}{2} (\eta L  + \eta' L + n_x n_q L )   n_y L  } e^{- \frac{qB}{4}  | (\eta-\eta') L \mathbf{ e}_x + \mathbf{ n}_B L   |^2  }   \nonumber \\
&     \times \Gamma(\frac{1}{2} - \frac{\mu \varepsilon}{q B})  U (\frac{1}{2} - \frac{\mu \varepsilon}{q B}, 1,  \frac{qB}{2}  | (\eta-\eta') L \mathbf{ e}_x  + \mathbf{ n}_B L   |^2  )   ,
\end{align}
where
\begin{align}
 & \mathcal{M}^{( \infty, 2D)}_B( \eta L \mathbf{ e}_x , \eta'  L\mathbf{ e}_x;  \varepsilon)   \nonumber \\
  & = - \frac{4}{2\mu}  G^{( \infty, 2D)}_B( \eta L \mathbf{ e}_x, \eta' L  \mathbf{ e}_x; \varepsilon) \nonumber \\
  & +  \delta_{\eta, \eta' }  \frac{1}{\pi} (\gamma_E + \frac{1}{2}\ln \frac{\mu \varepsilon \rho^2}{2}) |_{\rho \rightarrow 0} . 
\end{align}
Only diagonal terms of infinite volume magnetic Green's function, $ G^{( \infty, 2D)}_B  $, are  UV divergent. Using Eq.(\ref{kummerUV}) again, thus the  UV regularized diagonal terms of $\mathcal{M}^{( \infty, 2D)}_B$ function is   given again by
\begin{align}
  \mathcal{M}^{( \infty, 2D)}_B( \eta L \mathbf{ e}_x , \eta  L\mathbf{ e}_x;  \varepsilon)     =  -  \frac{1}{\pi} \left [  \psi(\frac{1}{2} - \frac{\mu \varepsilon}{qB}) + \ln \frac{ q B}{\mu \varepsilon} \right ]   . \label{zetaMinf2D}
\end{align}
The  example  plot of  $ \mathcal{M}^{( L,\frac{ \mathbf{ P}_B}{2} ,2D)}_B(\mathbf{ 0} , \mathbf{ 0}; \varepsilon) $  for various of $\mathbf{ P}_B$'s  compared with $ \mathcal{M}^{(   \infty ,2D)}_B(\mathbf{ 0} , \mathbf{ 0}; \varepsilon) $ is given in Fig.~\ref{Mplot}.

  \begin{figure}
\begin{center}
\includegraphics[width=1.0\textwidth]{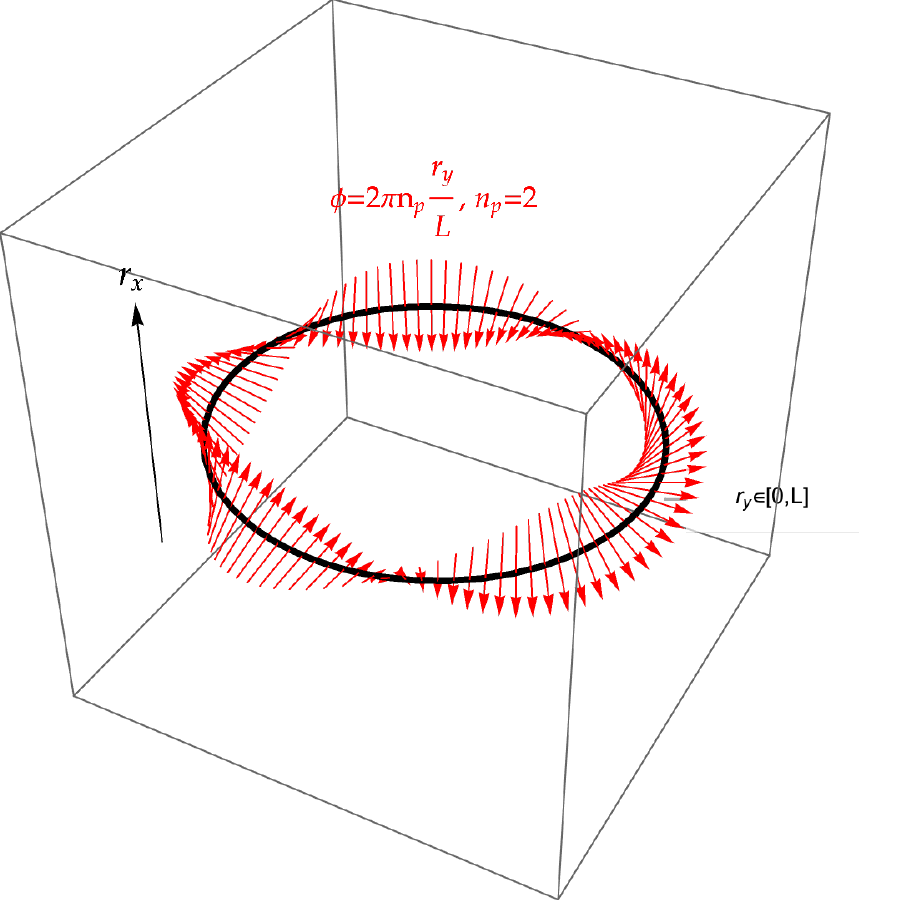}
\caption{  The plot of  rotation of   phase $ \varphi ( r_y)  $ defined in Eq.(\ref{phaserot})   along a cross section of a torus with a fixed $r_x$ (black circle). The phase  changes of $\varphi ( r_y) $ on the circle of $r_y$ is represented by the rotation of red arrows.  }\label{mobisplot}
\end{center}
\end{figure}

\section{Topological features of a magnetic system in a finite volume}\label{secberryphase}

It has been well-known   in condensed matter physics that the Bloch electron in a magnetic field yields a non-trivial topology \cite{RevModPhys.82.1959}. The non-trivial topology of a magnetic system can be visually illustrated  simply by using the twisted boundary condition given in Eq.(\ref{magneticperiodic}). Two edges of enlarged magnetic cubic box at $r_x=0$ and $r_x=n_q L$   are glued together by a   twist in the phase of wavefunction: 
\begin{equation}
\psi^{(L, \frac{\mathbf{ P}_B}{2})}_\varepsilon( n_q L, r_y )= e^{ i \frac{\mathbf{ P}_B}{2} \cdot \mathbf{ n}_B L}   e^{-i  \varphi ( r_y) } \psi^{(L, \frac{\mathbf{ P}_B}{2})}_\varepsilon( 0, r_y ) ,\label{twistbc}
\end{equation}
where 
\begin{equation}
\varphi ( r_y) = qB n_q L r_y = 2\pi n_p \frac{r_y}{L} \label{phaserot}
\end{equation}
is  the twisted phase  of wavefunction   along the circle of $r_y \in [0,L]$ that is a cross section of a torus with a fixed $r_x$. How much of twists in the phase   is totally determined by   $ n_p$.  Hence, the phase rotation of $\varphi ( r_y)$ along  the circle of $r_y$ in fact  form a geometry of $n_p$ times twisted M\"obius strip, see  an example in Fig.\ref{mobisplot}, which demonstrates a non-trivial topology.

\subsection{$\mathbf{ k}$-space and Brillouin zone}

Although in LQCD, the parameter $ \mathbf{ P}_B $ is associated with the plane wave of  CM motion of two-particle system, $e^{i \mathbf{ P}_B \cdot \mathbf{ R}}$, see Ref.~\cite{Guo:2021lhz}. Requirement of periodic boundary condition in CM motion yields the discrete value of   $ \mathbf{ P}_B $'s in Eq.(\ref{PBvecdiscrete}), and   discrete energy spectra as well.
To further examine some non-trivial topological features and analytic  properties of a magnetic system in finite volume, from now on, the discrete magnetic lattice vector $\frac{\mathbf{ P}_B}{2}$ is replaced by a continuous  wave vector $\mathbf{ k}$ that is analogous  to the crystal momentum in condensed matter physics. In current section and also Sec.~\ref{edgestates}, the wave vector  $\mathbf{ k}$ are   limited in real space.  The continuous distribution of  wave vector $\mathbf{ k}$ allows the introduction of Berry phase that is defined in a real $\mathbf{ k}$-space \cite{MVBerry03081984,RevModPhys.82.1959}. The Berry phase is a phase angle that describes the global phase evolution of the wavefunction of a system in a closed path in  $\mathbf{ k}$-space.  Due to the fact that  the same physical state is represented by  a ray of wavefunctions  that differ by a phase, such as $|\psi \rangle $ and $|\psi' \rangle =e^{i \phi }|\psi \rangle $,  the set of phase factor $e^{i \phi }$  form a $U(1)$ group.  Hence the ray of wavefunctions that are connected by a phase factor    define a $U(1)$ fibre in a manifold of $\mathbf{ k}$-space. Therefore, Berry phase is also recognized as a topological holonomy of the connection defined in a $U(1)$   fibre bundle in a parameter space \cite{PhysRevLett.51.2167}, which is $\mathbf{ k}$-space in our case. Berry phase is   an important physical quantity that measures the topological feature of a system in a parameter space.

  When $\mathbf{ k}$ is varied continuously,  the discrete energy spectra  are   smeared into  energy bands, also called bulk energy bands. These energy bands are separated by forbidden gaps between them due to the particle interactions. Each single allowed energy band hence becomes an isolated island in totally periodic systems. It has been known that the  edge effects   in a non-trivial topological system may allow the gapless energy solutions \cite{PhysRevLett.71.3697,PhysRevB.48.11851}, which yields a continuous and smooth connection between  two isolated energy bands. The topological edge solutions in a magnetic system will be discussed in Sec.~\ref{edgestates}. In addition, when  wave vector $\mathbf{ k}$ is further extended into a complex plane, given certain paths, the real energy solutions in the gap can also be found, which also connect two isolated energy bands smoothly. The discussion of analytic continuation of solutions in forbidden gaps  will be presented in Sec.~\ref{analyticproperty}.

Using Eq.(\ref{magneticGreendef}), one can show easily  that
\begin{equation}
 G^{(L, \mathbf{ k} +\mathbf{ G} ,2D )}_B(\bm{\rho}, \bm{\rho}' ; \varepsilon)  =  G^{(L, \mathbf{ k}  )}_B(\bm{\rho}, \bm{\rho}' ; \varepsilon)   ,  
\end{equation}
where
\begin{equation}
\mathbf{ G} =  \frac{2\pi}{n_q L} \mathbf{ e}_x + \frac{2\pi}{  L} \mathbf{ e}_y,
\end{equation}
hence $ \psi^{(L,   \mathbf{ k}+ \mathbf{ G} ,2D )}_{  \varepsilon}(\mathbf{ r})$   satisfies LS equation
\begin{align}
&  \psi^{(L,   \mathbf{ k}+\mathbf{ G} , 2D )}_{  \varepsilon}(\bm{\rho})  \nonumber \\
&= \int_{L_B^2} d \bm{\rho}' G^{(L, \mathbf{ k})}_B( \bm{\rho}, \bm{\rho}' ; \varepsilon) V^{(L)}(\bm{\rho}')\psi^{(L, \mathbf{ k} +  \mathbf{ G}, 2D )}_{  \varepsilon}(\bm{\rho}')  ,  
\end{align}
so does $ \psi^{(L,   \mathbf{ k} ,2D )}_{  \varepsilon}(\bm{\rho})$.  Therefore $ \psi^{(L,   \mathbf{ k}+ \mathbf{ G} , 2D   )}_{  \varepsilon}(\bm{\rho})$ and $ \psi^{(L,   \mathbf{ k} ,2D  )}_{  \varepsilon}(\bm{\rho})$ can only differ by a arbitrary phase factor, such as,
\begin{equation}
 \psi^{(L,   \mathbf{ k}+ \mathbf{ G} , 2D   )}_{  \varepsilon}(\bm{\rho}) =  \psi^{(L,   \mathbf{ k} ,2D  )}_{  \varepsilon}(\bm{\rho}),
\end{equation}
and  they both describe the same physical state. Therefore $  \mathbf{ k}+ \mathbf{ G} $ and $  \mathbf{ k} $ are identified as the same point. The wave vector $\mathbf{ k}$ hence can be limited in first magnetic Brillouin zone,
\begin{equation}
k_x \in [0, \frac{2\pi}{n_q L}], \ \ \ \   k_y \in [0, \frac{2\pi}{L}],
\end{equation}
 and the entire Brillouin zone form the geometry of a torus.

\subsection{Berry phase and Berry vector potential}

The non-trivial topology of magnetic system in finite volume results in a non-zero Berry phase. The Berry phase  is defined crossing over the torus of entire Brillouin zone by
\begin{align}
\gamma_\varepsilon  &=  \int_{0}^{ \frac{2\pi}{  L}} d k_y \left [  A_{\varepsilon,y} ( \frac{2\pi}{n_q L} , k_y  ) - A_{\varepsilon,y} ( 0 , k_y  )  \right ] \nonumber \\
& -   \int_{0}^{ \frac{2\pi}{n_q L}} d k_x \left [  A_{\varepsilon,x} ( k_x, \frac{2\pi}{L}   ) - A_{\varepsilon,x} (  k_x,0  )  \right ], \label{Berryphase}
\end{align}
where Berry vector potential $\mathbf{ A} (k_x, k_y)$ is given by
\begin{equation}
\mathbf{ A}_\varepsilon (k_x, k_y) =  \int_{L_B^2} d \bm{\rho}  u^{(   \mathbf{ k} ,2D  )*}_{  \varepsilon}(\bm{\rho}) i  \nabla_{\mathbf{ k}}   u^{(   \mathbf{ k} ,2D  )}_{  \varepsilon}(\bm{\rho}),
\end{equation}
and $ u^{(  \mathbf{ k}  ,2D )}_{  \varepsilon}(\bm{\rho})$ stands for the Bloch wavefunction and is related to $\psi^{(L,   \mathbf{ k}  ,2D )}_{  \varepsilon}(\bm{\rho})$ by
\begin{equation}
 \psi^{(L,   \mathbf{ k},2D  )}_{  \varepsilon}(\bm{\rho}) = e^{i \mathbf{ k} \cdot \bm{\rho}} u^{(   \mathbf{ k},2D   )}_{  \varepsilon}(\bm{\rho}).
\end{equation}
The Berry phase over  the torus of entire Brillouin zone is in fact a topological invariant quantity and quantized as $2\pi$ multiplied by an integer that is known as a Chern number  \cite{PhysRevLett.51.2167}.

In general, the Berry phase must be computed numerically by solving eigenvalue problems.  In presence of particles interactions, the wavefunction must be given by linear superposition of 
\begin{equation}
 \psi^{(L,   \mathbf{ k},2D  )}_{  \varepsilon}(\bm{\rho})= \sum_{n =0}^\infty  \sum_{\alpha=0}^{n_p -1}    c_{n,\alpha}^{(  \mathbf{ k})} (\varepsilon) \chi_{n,\alpha}^{(  \mathbf{ k})}(\bm{\rho} )  .
\end{equation}
The coefficient $c_{n,\alpha}^{(  \mathbf{ k})} (\varepsilon)$ satisfies a matrix equation,
 \begin{equation}
H^{( \mathbf{ k})} c^{( \mathbf{ k})}  (\varepsilon) =       \varepsilon    c^{( \mathbf{ k})}  (\varepsilon)  , \label{Heff}
\end{equation}
where  the matrix elements of effective Hamiltonian $H^{( \mathbf{ k})}$ are  given by
\begin{equation}
H^{( \mathbf{ k})}_{n, \alpha; n', \alpha'} = \delta_{n,\alpha; n', \alpha'}    \frac{qB}{\mu}(n+ \frac{1}{2})    + V^{( \mathbf{ k})}_{n,\alpha; n', \alpha'} ,
\end{equation}
and $V^{( \mathbf{ k})}_{n,\alpha; n', \alpha'} $ is defined in Eq.(\ref{Vmatrixelement}). The wave vector $ \mathbf{ k}$ in Eq.(\ref{Heff}) is now treated as the parameter of dynamics of system, and ultimately, adiabatic evolution of $ \mathbf{ k}$ crossing over magnetic Brillouin zone yields a Berry phase  \cite{RevModPhys.82.1959}.

Since Berry phase is a topological invariance and also a robust quantity against particle interactions, non-trivial topological feature  of a magnetic system in finite volume can be demonstrated by only using solutions of zero particle interactions. For a fixed $n$, there are $n_p$ degenerate states, 
\begin{equation}
u_{n,\alpha}^{( \mathbf{ k})}(\bm{\rho} ) =e^{-i \mathbf{ k} \cdot \bm{\rho}}  \chi_{n,\alpha}^{( \mathbf{ k})}(\bm{\rho} ) .
\end{equation}
The Berry phase for degenerate states $u_{n,\alpha}^{( \mathbf{ k})}(\bm{\rho} ) $with a fixed $n$ is defined by the trace of Berry phase for each state \cite{vanderbilt_2018},
\begin{equation}
\gamma_n = \sum_{\alpha =0}^{n_p -1} \gamma_{n, \alpha},
\end{equation}
 where $\gamma_{n, \alpha}$ is defined in Eq.(\ref{Berryphase}) with Berry vector potential,
\begin{equation}
\mathbf{ A}_{n,\alpha} (k_x, k_y) =  \int_{L_B^2} d \bm{\rho}  u^{(   \mathbf{ k}    )*}_{  n,\alpha}(\bm{\rho}) i  \nabla_{\mathbf{ k}}   u^{(   \mathbf{ k}   )}_{  n , \alpha}(\bm{\rho}).
\end{equation}

Using relations of 1D harmonic oscillator eigen-solutions,
 \begin{align}
& \sqrt{qB} r_x \phi_n (r_x) =\sqrt{ \frac{ n+1}{2} }\phi_{n+1} (r_x) + \sqrt{\frac{n}{2} } \phi_{n-1} (r_x) , \nonumber \\
 &   \partial_{\sqrt{ q B} r_x}  \phi_n(r_x)    =  -  \sqrt{ \frac{ n+1}{2} } \phi_{n+1} (r_x) + \sqrt{\frac{n}{2} } \phi_{n-1} (r_x)      ,
 \end{align}
we find
 \begin{align}
& i \partial_{k_x} u_{n,\alpha}^{( \mathbf{ k})}(\bm{\rho} ) \nonumber \\
& = \frac{1}{\sqrt{qB}}  \left [ \sqrt{ \frac{ n+1}{2} } u_{n+1,\alpha}^{( \mathbf{ k})}(\bm{\rho} ) + \sqrt{\frac{n}{2} }  u_{n-1,\alpha}^{( \mathbf{ k})}(\bm{\rho} )  \right ] \nonumber \\
 &     -   (  \frac{ \frac{2\pi    \alpha}{L} + k_y }{qB})   u_{n,\alpha}^{( \mathbf{ k})}(\bm{\rho} ) , \nonumber \\
& i  \partial_{k_y} u_{n,\alpha}^{( \mathbf{ k})}(\bm{\rho} )    \nonumber \\
&    = - \frac{i}{\sqrt{qB}}  \left [  \sqrt{ \frac{ n+1}{2} } u_{n+1,\alpha}^{( \mathbf{ k})}(\bm{\rho} )- \sqrt{\frac{n}{2} }   u_{n-1,\alpha}^{( \mathbf{ k})}(\bm{\rho} )     \right ]. \label{diffuneq}
 \end{align}
Also using orthogonality relation given in Eq.(\ref{chiorthogonal}),   we thus obtain
   \begin{align}
  A_{n, \alpha ,x} (  k_x, k_y  ) &=    -   (  \frac{ \frac{2\pi    \alpha}{L} + k_y }{qB}) ,  \nonumber \\
  A_{n, \alpha ,y} (  k_x, k_y  )  &=  0 . \label{berrryvecpot}
 \end{align}
 Hence, the Berry phases are given by
 \begin{equation}
 \gamma_{n, \alpha} = \frac{2\pi}{n_p}, \ \ \ \  \frac{\gamma_n}{2\pi} =1. \label{berryphase}
 \end{equation}

\subsection{Topological properties of $\chi_{n,\alpha}^{( \mathbf{ k}    )}$ functions} 
The Berry phase of a magnetic system in finite volume can also be understood by simply examining the topological properties of $\chi_{n,\alpha}^{( \mathbf{ k}    )}$ functions.

 Using Eq.(\ref{chidef}), one can show that  how the center of  $\chi_{n,\alpha}^{( \mathbf{ k}    )}$ function is pushed along $x$-direction when the wave vector $\mathbf{ k} $ is forced to change in $y$-direction, 
   \begin{equation}
  \chi_{n,\alpha}^{( \mathbf{ k}  + \triangle k_y \mathbf{ e}_y )}(\bm{\rho} ) = e^{ i \triangle k_y  r_y}  \chi_{n,\alpha}^{( \mathbf{ k}    )}(\bm{\rho}  +  \frac{ \triangle k_y   }{qB} \mathbf{ e}_x) . \label{chidky}
 \end{equation}
 Thus  Eq.(\ref{chidky}) yields
  \begin{equation}
  \chi_{n,\alpha}^{( \mathbf{ k}  + \frac{2\pi}{L} \frac{n_p}{n_q}   \mathbf{ e}_y )}(\bm{\rho} )  
   = e^{ i  q B  L r_y}  \chi_{n,\alpha}^{( \mathbf{ k}    )}(\bm{\rho}  + L   \mathbf{ e}_x),
 \end{equation}
 that is to say, to move the center of  $\chi_{n,\alpha}^{( \mathbf{ k}    )}$  by length-$L$ in $x$-direction, it requires the change of wave vector $\mathbf{ k}$ by  $\frac{2\pi}{L} \frac{n_p}{n_q} $ in y-direction.

When  $\mathbf{ k} $ is forced to move across entire Brillouin zone in $y$-direction,  
   \begin{equation}
  \chi_{n,\alpha}^{( \mathbf{ k}  + \frac{2\pi}{L} \mathbf{ e}_y )}(\bm{\rho} )  
   = e^{ i  \frac{2\pi}{L}  r_y}  \chi_{n,\alpha}^{( \mathbf{ k}    )}(\bm{\rho}  +  \frac{  n_q L  }{n_p} \mathbf{ e}_x)  =  \chi_{n,\alpha+1}^{( \mathbf{ k}    )}(\bm{\rho} ) , 
    \end{equation}
  the center of  $\chi_{n,\alpha}^{( \mathbf{ k}    )}$  function is only moved by $\frac{n_q L}{n_p}$ in $x$-direction, which  is then smoothly connected to the  $\chi_{n,\alpha+1}^{( \mathbf{ k}    )}$ function.   Hence,   an array of states 
\begin{equation}
\chi_{n}^{( \mathbf{ k}    )} =  \left [ \chi_{n,0}^{( \mathbf{ k}    )}, \chi_{n,1}^{( \mathbf{ k}    )}, \cdots, \chi_{n, n_p-1}^{( \mathbf{ k}    )} \right ],
\end{equation}
   behaves as a $n_p$ components spinor.  $\triangle k_y  = \frac{2\pi}{L}$ plays the role of     raising operator which change  each individual component of  spinor $\chi_{n}^{( \mathbf{ k}    )}$  by one unit,
   \begin{equation}
\chi_{n}^{( \mathbf{ k} +   \frac{2\pi}{L} \mathbf{ e}_y  )} =  \left [ \chi_{n,1}^{( \mathbf{ k}    )}, \chi_{n,2}^{( \mathbf{ k}    )}, \cdots, \chi_{n, n_p}^{( \mathbf{ k}    )} \right ].
\end{equation}
    Operating   raising operator $n_p$ times,   with the help of periodic boundary condition,    we can also show that 
   \begin{align}
    \chi_{n,\alpha}^{( \mathbf{ k}  + \frac{2\pi}{L} n_p \mathbf{ e}_y )}(\bm{\rho} )  
   &= e^{ i  \frac{2\pi}{L} n_p r_y}  \chi_{n,\alpha}^{( \mathbf{ k}    )}(\bm{\rho}  +  n_q L   \mathbf{ e}_x)  =  \chi_{n,\alpha+ n_p}^{( \mathbf{ k}    )}(\bm{\rho} ) \nonumber \\
   & = e^{i k_x n_q L}    \chi_{n,\alpha}^{( \mathbf{ k}    )}(\bm{\rho} )   . \label{chitwistedbd}
 \end{align}
 Therefore,  
    \begin{equation}
\chi_{n}^{( \mathbf{ k} +   \frac{2\pi}{L} \mathbf{ e}_y  )} =  \left [ \chi_{n,1}^{( \mathbf{ k}    )}, \chi_{n,2}^{( \mathbf{ k}    )}, \cdots, e^{i k_x n_q L} \chi_{n, 0}^{( \mathbf{ k}    )} \right ], \label{chispinorky}
\end{equation}
changing $\mathbf{ k}$ by $ \frac{2\pi}{L}   \mathbf{ e}_y$ leads to the circulation of all components only once, and only the component sitting at right edge of spinor gains a phase factor, $e^{i k_x n_q L}  $.
On the other hand, changing $\mathbf{ k}$ by $ \frac{2\pi}{L} n_p \mathbf{ e}_y$ however yields that the center of each component of spinor  $\chi_{n}^{( \mathbf{ k}    )}$  is   forced to wind around  entire magnetic unit cell in $x$-direction.  Meanwhile, all components  of the spinor  circulate $n_p$ times and come back to the starting point, so each one has a  chance to gain a phase factor  $e^{i k_x n_q L}  $ when it reaches the right edge of spinor,
    \begin{equation}
\chi_{n}^{( \mathbf{ k} +   \frac{2\pi}{L} n_p \mathbf{ e}_y  )} = e^{i k_x n_q L}  \chi_{n}^{( \mathbf{ k}  )} .
\end{equation}

  In addition, when the  wave vector $\mathbf{ k}$ is forced to move across magnetic Brillouin zone in $x$-direction, the each component of spinor $\chi_{n}^{( \mathbf{ k}    )}$  remains at the same location in a spinor,
    \begin{equation}
  \chi_{n,\alpha}^{( \mathbf{ k}  + \frac{2\pi}{n_q L} \mathbf{ e}_x )}(\bm{\rho} )  
     =  \chi_{n,\alpha}^{( \mathbf{ k}    )}(\bm{\rho} )  ,
 \end{equation}
 and
    \begin{equation}
\chi_{n}^{( \mathbf{ k} +   \frac{2\pi}{n_q L} \mathbf{ e}_x  )} = \chi_{n}^{( \mathbf{ k} )}   . \label{chispinorkx}
\end{equation}

 Now, the non-trivial Berry phase may also be understood simply by using the   properties given in Eq.(\ref{chispinorky}) and Eq.(\ref{chispinorkx}).   Assuming  that we start at one corner of Brillouin zone: $\mathbf{ k}= (0,0)$ with initial spinor of
     \begin{equation}
  \chi_{n}^{(  i  )} = \left [ \chi_{n,0}^{(  0,0    )}, \chi_{n,1}^{( 0,0  )}, \cdots, \chi_{n, n_p-1}^{( 0,0   )} \right ] ,
\end{equation}
where $i$ is used to label initial state of spinor, 
    then we start moving around the boundary of  magnetic Brillouin zone counter-clock wise,
 \begin{equation}
 \mathbf{ k}:  (0,0) \stackrel{(1)}{ \rightarrow } ( \frac{2\pi}{n_q L},0) \stackrel{(2)}{ \rightarrow } ( \frac{2\pi}{n_q L}, \frac{2\pi}{L})\stackrel{(3)}{ \rightarrow } ( 0, \frac{2\pi}{L})\stackrel{(4)}{ \rightarrow } ( 0, 0).
 \end{equation}

At step (1), moving from $ \mathbf{ k}=  (0,0)$ to $( \frac{2\pi}{n_q L},0)$ by  an  increase of   $\triangle k_x= \frac{2\pi}{n_q L}$, there is no phase change,
    \begin{equation}
\chi_{n}^{(     \frac{2\pi}{n_q L}   , 0 )} =  \chi_{n}^{(  i  )}  . 
\end{equation}

At step (2), moving from $( \frac{2\pi}{n_q L},0)$ to $( \frac{2\pi}{n_q L}, \frac{2\pi}{L})$  by an increase of  $\triangle k_y= \frac{2\pi}{  L}$, we find
     \begin{equation}
\chi_{n}^{(\frac{2\pi}{n_q L}, \frac{2\pi}{L} )} =  \left [ \chi_{n,1}^{( 0,0   )}, \chi_{n,2}^{(0,0 )}, \cdots, e^{i \frac{2\pi}{n_q L} n_q L} \chi_{n, 0}^{(0,0   )} \right ].
\end{equation}

 At step (3), moving from $( \frac{2\pi}{n_q L}, \frac{2\pi}{L})$  to $(0,  \frac{2\pi}{ L})$ by  a decrease fo    $\triangle k_x= -\frac{2\pi}{n_q L}$, there is again no phase change, so that
    \begin{equation}
\chi_{n}^{( 0, \frac{2\pi}{L} )} =\chi_{n}^{(\frac{2\pi}{n_q L}, \frac{2\pi}{L} )} . 
\end{equation}

At last step (4),  moving from   $(0,  \frac{2\pi}{ L})$  back to $(  0,0)$  by a decrease of   $\triangle k_y= -\frac{2\pi}{  L}$,  although   there is no phase change at last step, all components of spinor are moved down by one unit, so that the final state of spinor is given by
     \begin{equation}
\chi_{n}^{( f)} =  \left [ \chi_{n,0}^{( 0,0   )}, \chi_{n,1}^{(0,0 )}, \cdots, e^{i \frac{2\pi}{n_q L} n_q L} \chi_{n, n_p-1}^{(0,0   )} \right ].
\end{equation}
  Therefore, the phase difference between initial and final states is given by
 \begin{equation}
- \sum_{\alpha=0}^{n_p-1} Im  \ln \langle \chi_{n, \alpha}^{( f)}  | \chi_{n, \alpha}^{( i)}  \rangle =   \frac{2\pi}{n_q L} n_q L = 2\pi,
 \end{equation}
which   can   be identified as Berry phase $\gamma_n$.

  \begin{figure}
\begin{center}
\includegraphics[width=1.0\textwidth]{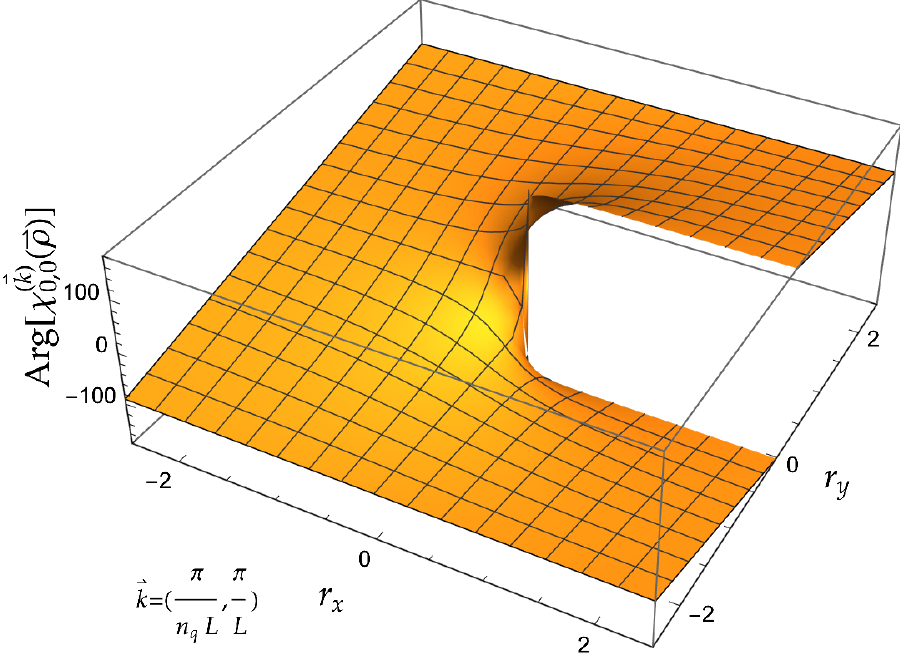}
\includegraphics[width=1.0\textwidth]{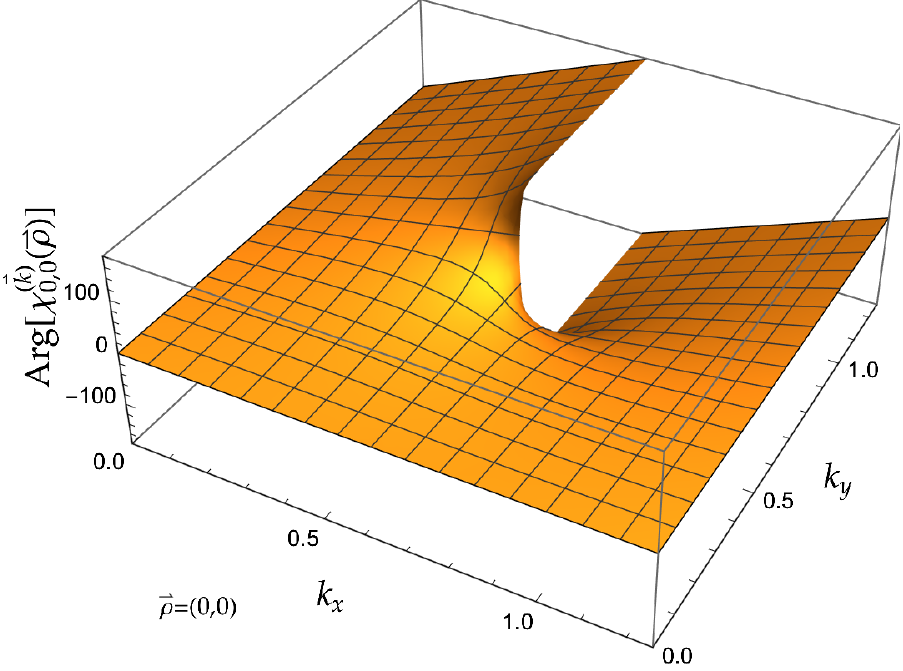}
\caption{  The phase plot of $\chi_{0,0}^{( \mathbf{ k})}(\bm{\rho} )  $ with fixed $\mathbf{ k}= (  \frac{\pi}{L},  \frac{\pi}{L} )$ in $\bm{\rho} $-space (upper pannel) vs. phase plot of $\chi_{0,0}^{( \mathbf{ k})}(\bm{\rho} )  $ with fixed $\bm{\rho} = (0,0)$ in $\mathbf{ k}$-space (lower panel). The parameters are chosen as: $L=5$ and $n_p=n_q=1$.}\label{xnphaseplot}
\end{center}
\end{figure}

 The Berry phase is the quantity that describes the accumulation of a global phase of a  system's wavefunction as the $\mathbf{ k}$ is carried around the torus of   Brillouin zone,  non-zero value of Berry  phase hence represents a topological obstruction to the determination  of the phase of wavefunction \cite{KOHMOTO1985343} over entire Brillouin zone.  For a magnetic system, the magnetic field create a vortex-like singularities in wavefunctions that attribute to a non-trivial topology of a magnetic system. The vortex-like singularities can be illustrated analytically by $ \chi_{0, \alpha}^{( \mathbf{ k} )} (\bm{\rho})$. Using Eq.(\ref{chidef}) and $H_0 (x)=1$, we find  
   \begin{align}
  &  \Theta_{\alpha}^{( \mathbf{ k})}(\bm{\rho} )   \nonumber \\
  & = e^{-i ( k_y+\frac{2\pi   \alpha }{L}  )r_y  }  \chi_{0,\alpha}^{( \mathbf{ k})}(\bm{\rho} )  
 \nonumber \\
 & =      \frac{1}{\sqrt{L}} \left ( \frac{qB}{\pi} \right )^{\frac{1}{4}} e^{- \frac{q B}{2} (r_x  + \frac{   k_y+\frac{2\pi  \alpha}{L}  }{qB}  )^2  }  \nonumber \\
  & \times   \vartheta_3 \left (\frac{\pi n_p}{L} \left [  ( r_y-\frac{k_x}{qB}  ) + i (r_x  + \frac{   k_y+\frac{2\pi  \alpha}{L}  }{qB}       ) \right  ] , e^{- \pi n_p n_q  } \right )    , \label{thetafunc}
 \end{align}
 where  $  \vartheta_3 (z, q)$ is Jacobi's theta function  \cite{NIST:DLMF}, and defined by 
 \begin{equation}
   \vartheta_3  (z,q) =1+2 \sum_{n=1}^\infty q^{n^2} \cos (2 n z).
 \end{equation}
 The zeros of $$  \vartheta_3 (z,  q= e^{i \pi \tau} )$$ are determined by linear equation,
 \begin{equation}
 z=( n_1 + \frac{1}{2}) \pi +( n_2  + \frac{1}{2}) \tau \pi, \ \ \ \  (n_1, n_2) \in \mathbb{Z},
 \end{equation}
  hence, the locations of zeros of $\chi_{0,\alpha}^{( \mathbf{ k})}(\bm{\rho} )  $ are  given by 
       \begin{align}
  &     r_y-\frac{k_x}{qB}      =( n_1 + \frac{1}{2}) \frac{L}{n_p}, \ \ n_1 \in \mathbb{Z},  \nonumber \\
  &    r_x  +  \frac{   k_y }{qB} + \frac{ \frac{2\pi  \alpha}{L}  }{qB}      = (   n_2       + \frac{   1 }{2}  ) n_q L, \ \ n_2 \in \mathbb{Z}.
 \end{align}
 The zeros of $\chi_{0,\alpha}^{( \mathbf{ k})}(\bm{\rho} )  $ present vortex-like singularities, which ultimately create discontinuity of phase of   $\chi_{0,\alpha}^{( \mathbf{ k})}(\bm{\rho} )  $ in both $\bm{\rho}$- and $\mathbf{ k}$-space.  The  $  \vartheta_3 (z, q)$ is a real function when   $z $ values are real and $|q|<1$, therefore, for a fixed $\mathbf{ k}$,
 \begin{equation}
Im \left [ \Theta_{\alpha}^{( \mathbf{ k})}(\bm{\rho} ) \right ]_{\bm{\rho} = ( - \frac{   k_y+\frac{2\pi  \alpha}{L}  }{qB}  , r_y)}   =0,
 \end{equation}
thus the phase of $\chi_{0,\alpha}^{( \mathbf{ k})}(\bm{\rho} )  $ is not well-defined along the line of $  ( - \frac{   k_y+\frac{2\pi  \alpha}{L}  }{qB}  , r_y)$  in $\bm{\rho}$-space. On the other hand, for a fixed $\bm{\rho}$, 
   \begin{equation}
Im \left [ \Theta_{\alpha}^{( \mathbf{ k})}(\bm{\rho} ) \right ]_{ \mathbf{ k} = ( k_x, -  \frac{2\pi  \alpha}{L}   -q B r_y)}   =0,
 \end{equation}
  so in $\mathbf{ k}$-space, the phase is also not well-defined along the line of $   ( k_x, -  \frac{2\pi  \alpha}{L}   -q B r_y)$. These two lines cut though both entire $\bm{\rho}$- and $\mathbf{ k}$-space. Because of asymmetry of $\Theta_{\alpha}^{( \mathbf{ k})}(\bm{\rho} ) $ along these two lines, it creates the mismatch of the phase of $\chi_{0,\alpha}^{( \mathbf{ k})}(\bm{\rho} )  $ on half portion of the line,  which starts at the location of zeros of $\chi_{0,\alpha}^{( \mathbf{ k})}(\bm{\rho} )  $, see Fig.~\ref{xnphaseplot} as an example of phase mismatch. These vortex-like singularities in wavefunction is similar to the branch point singularities in complex analysis, the vortex creates a cut in both $\bm{\rho}$- and $\mathbf{ k}$-space, and phase of wavefunction along the cut has a discontinuity. Hence, when particle is forced to wind around the vortex, the phase of wavefunction has a jump which account how many times the winding number of motion around the vortex.

  \begin{figure}
\begin{center}
\includegraphics[width=1.0\textwidth]{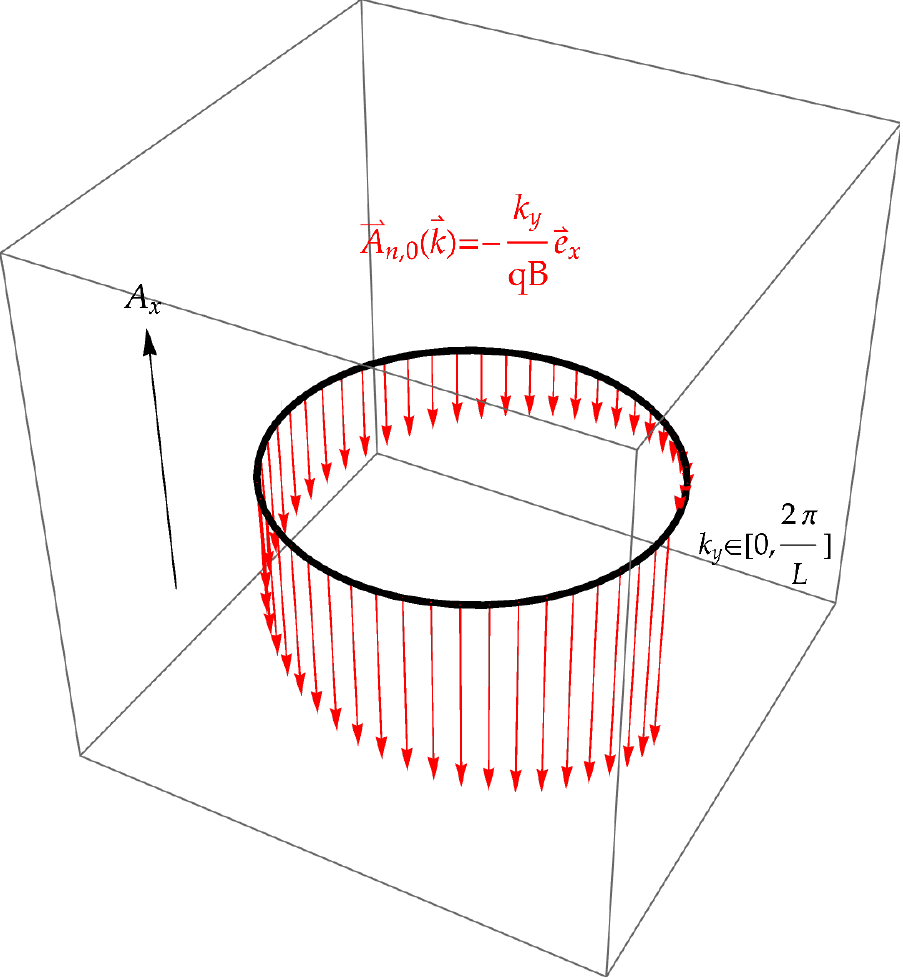}
\caption{  The   plot of Berry vector potential $\mathbf{ A}_{n,0} (k_x,k_y) $  on a cross section of torus of Brillouin zone with a fixed $k_x$. The parameters are chosen as: $L=5$ and $n_p=n_q=1$.}\label{berryvecplot}
\end{center}
\end{figure}

The phase discontinuity of $\chi_{n,\alpha}^{( \mathbf{ k})}(\bm{\rho} )  $ ultimately creates non-trivial topology of the Berry vector potential given in Eq.(\ref{berrryvecpot}). The vortex-like singularities not only create discontinuity of phase in wavefunction, but also leads to the mismatch of Berry vector potential on the torus of entire magnetic Brillouin zone. Since the torus has no boundary, uniquely and smoothly defined Berry vector potential on the torus results in the trivial topology and vanishing Berry phase. For $\chi_{n,\alpha}^{( \mathbf{ k})}(\bm{\rho} )  $ wavefunction, according to Eq.(\ref{berrryvecpot}), it is clearly that the Berry vector potential  on the lower edge of torus along the line $\mathbf{ k} = ( k_x,0)$ is 
  \begin{equation}
  \mathbf{ A}_{n, \alpha } (  k_x,0 ) =    -         \alpha \frac{n_q L}{n_p}   \mathbf{ e}_x .
 \end{equation}
On the upper edge of torus along the line of  $\mathbf{ k} = ( k_x,\frac{2\pi}{L})$, the Berry vector potential is 
  \begin{equation}
  \mathbf{ A}_{n, \alpha } (   k_x,\frac{2\pi}{L} ) =    -        ( \alpha+1) \frac{n_q L}{n_p}   \mathbf{ e}_x .
 \end{equation}
The upper and lower edges of a torus is considered as the same points, hence, magnetic field ultimately cause a mismatch of Berry vector potential on the torus,  see    Fig.~\ref{berryvecplot} as an example. The discontinuity of Berry vector potential is given by
  \begin{equation}
   \mathbf{ A}_{n, \alpha } (  k_x, \frac{2\pi}{L} ) -   \mathbf{ A}_{n, \alpha } (  k_x  ,0  )     = - \frac{n_q L}{n_p}  \mathbf{ e}_x,
 \end{equation}
 which   ultimately leads to a non-zero Berry phase. 
 The discontinuity of Berry vector potential   on a closed path in $\mathbf{ k}$-space is known as a  holonomy \cite{PhysRevLett.51.2167}.  
 When wave vector $\mathbf{ k}$ is forced to move along a closed path, the Berry vector potential then  generates a horizontal  lift of the wavefunction    along the $U(1)$ fibre of each state, hence, in adiabatic limit,  the states along the path in $\mathbf{ k}$-space are all connected by 
 \begin{equation}
u_{n,\alpha}^{( \mathbf{ k} (t))}(\bm{\rho} )   \sim  e^{ -i \int_{\mathbf{ k} (0)}^{\mathbf{ k} (t)}  \mathbf{ A}_{n, \alpha } (\mathbf{ k})  \cdot d \mathbf{ k} }  u_{n,\alpha}^{( \mathbf{ k} (0))}(\bm{\rho} )  .
 \end{equation}
 Each state on the path has the memory of previous states along the path. The holonomy of a   system detects a topological or geometric nature of the underlying structure of the physical system. The twisting of $U(1)$ fibre bundle results in the non-trivial value of holonomy. The twisting of $U(1)$ fibre bundle in $\mathbf{ k}$-space can be understood by the relation given in Eq.(\ref{chitwistedbd}), which yields  
    \begin{equation}
   u_{n,\alpha}^{(  k_x ,   \frac{2\pi}{L} n_p   )}( \bm{\rho}  )     =   e^{ -i   \frac{2\pi}{L} n_p  r_y}  e^{i   \widetilde{\varphi} (k_x)  } u_{n,\alpha}^{(  k_x,  0   )}(     \bm{\rho}  ) , \label{untwistedbd}
 \end{equation} 
   where
 \begin{equation}
\widetilde{ \varphi}(k_x) = k_x n_q L.
 \end{equation}
 Eq.(\ref{untwistedbd}) may be interpreted as twisted boundary condition in enlarged Brillouin zone: 
 \begin{equation}
 k_x \in [0, \frac{2\pi}{ n_q L}], \ \ \ \  k_y \in   [0, \frac{2\pi}{   L} n_p].
 \end{equation}
 Hence, similar to twisted boundary condition in Eq.(\ref{twistbc}) in $\bm{\rho}$-space,  when two edges at $k_y =0 $ and $k_y = \frac{2\pi}{L} n_p$ of  enlarged Brillouin zone are glued together, $\widetilde{ \varphi}(k_x) $ describes the twisted phase of wavefunction along the circle of $k_x \in [ 0,  \frac{2\pi}{ n_q L}]$.

 We also remark that noticing  that Eq.(\ref{diffuneq}) may be rearranged to
 \begin{equation}
 \sum_{n', \alpha'}\left [ \delta_{n, \alpha; n', \alpha'} \nabla_{\mathbf{ k}} + i \mathbf{ A}_{n, \alpha; n',\alpha'} (\mathbf{ k} ) \right ] u_{n',\alpha'}^{( \mathbf{ k})}(\bm{\rho} ) =0,
 \end{equation}
where the matrix elements of Berry vector potential matrix are given by
 \begin{align}
   \mathbf{ A}_{n ,\alpha; n', \alpha'  } (  \mathbf{ k}   ) &=  \int_{L_B^2} d \bm{\rho}  u^{(   \mathbf{ k}    )*}_{  n,\alpha}(\bm{\rho}) i  \nabla_{\mathbf{ k}}   u^{(   \mathbf{ k}   )}_{  n' , \alpha'}(\bm{\rho})  \nonumber \\
  & = \frac{1}{\sqrt{qB}}    \sqrt{ \frac{ n+1}{2} } \delta_{n+1, n'}   \delta_{\alpha, \alpha'} (\mathbf{ e}_x  - i \mathbf{ e}_y) \nonumber \\
&+\frac{1}{\sqrt{qB}}    \sqrt{\frac{n}{2} }  \delta_{n-1, n'}  \delta_{\alpha, \alpha'} (\mathbf{ e}_x  + i \mathbf{ e}_y)    \nonumber \\
  &    -   \delta_{n, n'}  \delta_{\alpha, \alpha'}(  \frac{ \frac{2\pi    \alpha}{L} + k_y }{qB})  \mathbf{ e}_x.  
 \end{align}
Non-vanishing off-diagonal terms in Berry vector potential matrix suggest that a magnetic system may experience non-adiabatic transition between different eigen-states. For an example, assuming an non-interacting magnetic system, the general Bloch wavefunction is given by the linear superposition of eigen-states of non-interacting magnetic system,
\begin{equation}
u^{( \mathbf{ k} (t))}(\bm{\rho} ) = \sum_{n, \alpha } c_{n,\alpha} ( t) u_{n,\alpha}^{( \mathbf{ k} (t))}(\bm{\rho} ) ,
\end{equation}
where $t$ is used to parameterize the evolution of wave vector $\mathbf{ k}$. Also assuming $u^{( \mathbf{ k} (t))}(\bm{\rho} ) $ satisfies Schr\"odinger equation
\begin{equation}
 i \partial_t u^{( \mathbf{ k} (t))}(\bm{\rho} )  = \hat{H}_{eff} (\mathbf{ k} (t)) u^{( \mathbf{ k} (t))}(\bm{\rho} ) ,
\end{equation}
 where $u_{n,\alpha}^{( \mathbf{ k} (t))}(\bm{\rho} ) $ are eigen-solutions of $\hat{H}_{eff} (\mathbf{ k} (t))=e^{-i \mathbf{ k} (t) \cdot \bm{\rho}} \hat{H}_{\bm{\rho}} e^{i \mathbf{ k} (t) \cdot \bm{\rho}}$,
 \begin{equation}
 \hat{H}_{eff} (\mathbf{ k} (t)) u_{n,\alpha}^{( \mathbf{ k} (t))}(\bm{\rho} )  = \frac{qB}{\mu} (n+ \frac{1}{2}) u_{n,\alpha}^{( \mathbf{ k} (t))}(\bm{\rho} ) ,
 \end{equation}
  thus, we find that the coefficient $c_{n,\alpha} ( t)$ must satisfy equation,
   \begin{align}
     i \frac{d c_{n,\alpha} ( t) }{d t}   & =   \frac{qB}{\mu} (n+ \frac{1}{2})  c_{n,\alpha} ( t)   \nonumber \\
& - \frac{d  \mathbf{ k} (t) }{d t} \cdot \sum_{n', \alpha' }      \mathbf{ A}_{n ,\alpha; n', \alpha'  } (  \mathbf{ k}   )        c_{n',\alpha'} ( t) .
 \end{align}
  Therefore, the diagonal term in Berry vector potential matrix yields the Berry phase in Eq.(\ref{berryphase}) in the limit of adiabatic process, the off-diagonal terms may describes the transition among different eigen-states when $\mathbf{ k}$ is forced to increase or decrease.

\section{Topological edge states}\label{edgestates}

One of important consequences of a non-trivial topological system is the existence of gapless topological edge states that occur in the energy gap between the bulk bands \cite{PhysRevLett.71.3697,PhysRevB.48.11851}.     The study of conventional edge or surface states in fact has a long history \cite{PhysRev.56.317,zbMATH03006237}, the boundary effect may cause the localization of state near the edge or surface of material. Though the energy spectrum of a system with a penetrable boundary may protrude into the gap between bulk bands, for  topologically  trivial systems, the eigen-energies  of an impenetrable wall on boundary are only situated on the edge of bulk energy bands. This fact can be illustrated with a 2D system with different boundary conditions. The 2D finite volume Green's function that satisfies periodic boundary conditions in both $x$ and $y$ directions is given by
\begin{equation}
G_0^{(L,\mathbf{ k},2D)} (\bm{\rho}; \varepsilon) = \frac{1}{L^2} \sum_{\mathbf{ p} = \frac{2\pi \mathbf{ n}}{L} + \mathbf{ k} , \mathbf{ n } \in \mathbb{Z}^2} \frac{e^{i \mathbf{ p} \cdot \bm{\rho}}}{\varepsilon - \frac{\mathbf{ p}^2}{2\mu}}, \label{GL2Dbulk0}
\end{equation}
compared with
\begin{align}
&G_0^{(h.w., k_y \mathbf{ e}_y ,2D)} (\bm{\rho}, \bm{\rho}' ; \varepsilon) = \frac{1}{L} \sum_{ p_y = \frac{2\pi  n_y}{L} +  k_y, n_y \in \mathbb{Z} }  e^{i p_y  (r_y- r'_y) } \nonumber \\
& \times 2\mu \frac{ \sin \sqrt{2\mu \varepsilon - p_y^2} ( r_{y>} - \frac{ L}{2})\sin \sqrt{2\mu \varepsilon - p_y^2} ( r_{y<} + \frac{ L}{2}) }{ \sqrt{2\mu \varepsilon - p_y^2} \sin \sqrt{2\mu \varepsilon - p_y^2} L }, \label{GL2Dedge0}
\end{align}
 which satisfies hard wall boundary condition in $x$-direction but still remains periodic in $y$-direction,
 \begin{equation}
 G_0^{(h.w., k_y \mathbf{ e}_y ,2D)} (  \bm{\rho}, \bm{\rho}' ;  \varepsilon)  |_{ \bm{\rho}, \bm{\rho}' = \pm \frac{L}{2} \mathbf{ e}_x}  = 0 .
 \end{equation}
Therefore, with a contact interaction, and using Eq.(\ref{GL2Dbulk0}) and identity
\begin{equation}
\frac{1}{L} \sum_{ p_x = \frac{2\pi  n_x}{L} +  k_x, n_x \in \mathbb{Z} } \frac{1}{q^2 - p_x^2} = \frac{ \cot \frac{q - k_x}{2 } L+\cot \frac{q + k_x}{2 } L}{4 q},
\end{equation}
  the bulk energy band solutions with periodic boundary condition along both directions  are determined by
\begin{equation}
\frac{1}{V_0} =  \frac{2\mu}{L} \sum_{ p_y = \frac{2\pi  n_y}{L} +  k_y }^{n_y \in \mathbb{Z} } \!\!\!\!   \frac{\cot \frac{\sqrt{2\mu \varepsilon - p_y^2} - k_x}{2} L+\cot \frac{\sqrt{2\mu \varepsilon - p_y^2} + k_x}{2} L}{ 4 \sqrt{2\mu \varepsilon - p_y^2}}.
\end{equation} 
 Using  Eq.(\ref{GL2Dedge0}), the edge solutions with hard wall boundary condition along $x$-direction are determined by
\begin{equation}
\frac{1}{V_0} =  \frac{2\mu}{L} \sum_{ p_y = \frac{2\pi  n_y}{L} +  k_y, n_y \in \mathbb{Z}  }    \frac{\cot \frac{\sqrt{2\mu \varepsilon - p_y^2} -  \frac{\pi}{L} }{2} L}{ 2 \sqrt{2\mu \varepsilon - p_y^2}}.
\end{equation} 
We can see clearly that  for a fixed $k_y$ and $V_0$, the edge solution is only part of bulk energy band solutions with special value of wave vector $k_x= \frac{\pi}{L}$, which indeed sit at the edge of bulk energy bands.

On the contrary, even with  impenetrable walls on the boundary, the topological edge states may appear in the gap between bulk energy bands. In this section, we will show the solutions of various boundary conditions for a 2D magnetic system, and discuss how the boundary condition may affect the spectrum of a magnetic system.

\subsection{Various boundary condition solutions of 2D magnetic Green's function}
In this subsection, we study  various boundary condition solutions of 2D magnetic Green's function in $x$-direction, but the boundary condition in $y$-direction remains periodic. Hence,  the   solutions with different boundary conditions all have the form of 
\begin{align}
& G_B^{(k_y \mathbf{ e}_y ,2D)} (\bm{\rho},\bm{\rho}'; \varepsilon)  = \frac{1}{L} \sum_{ p_y = \frac{2\pi  n_y}{L} +  k_y, n_y \in \mathbb{Z} }  e^{i p_y  (r_y- r'_y) } \nonumber \\
& \times  G_B^{(1D)} (r_x + \frac{p_y}{qB}, r'_x+ \frac{p_y}{qB} ; \varepsilon  ) ,
\end{align}
where $G_B^{(1D)}$ satisfies differential equation
\begin{equation}
\left (\varepsilon + \frac{\partial_{r_x}^2 }{2\mu} - \frac{(qB)^2}{2\mu} r_x^2 \right )  G_B^{(1D)} (r_x  , r'_x  ; \varepsilon  ) = \delta(r_x - r'_x).  \label{G1Ddiffeq}
\end{equation}
Before the boundary condition is implemented, Eq.(\ref{G1Ddiffeq}) is parabolic cylinder equation type \cite{NIST:DLMF}, the homogeneous parabolic cylinder equation has two independent solutions called parabolic cylinder functions \cite{NIST:DLMF}: $$U(-\frac{\mu \varepsilon}{qB},  \pm \sqrt{2qB } r_x).$$ Therefore in general, the solution of  $G_B^{(1D)}$ is given by 
\begin{align}
&G_B^{(1D)}  (r_x  , r'_x  ; \varepsilon  ) \nonumber \\
&  = \left [ a U(-\frac{\mu \varepsilon}{qB},    \sqrt{2qB } r_{x<}) + b U(-\frac{\mu \varepsilon}{qB},  - \sqrt{2qB } r_{x<}) \right ] \nonumber \\
&  \times \left [ c U(-\frac{\mu \varepsilon}{qB},    \sqrt{2qB } r_{x>}) + d U(-\frac{\mu \varepsilon}{qB},  - \sqrt{2qB } r_{x>}) \right ], \label{G1Dgeneralsol}
\end{align}
where $r_{x <}$ and $r_{x<}$ refer to the lesser and greater of $(r_x, r'_x)$ respectively.
All coefficients $(a,b,c,d)$ are determined by boundary conditions and discontinuity relation
\begin{equation}
\partial_{r_x} G_B^{(1D)}  (r_x  , r'_x  ; \varepsilon  )|_{r_x = r'_x -0}^{r_x = r'_x + 0} =2\mu. \label{dGB1D}
\end{equation}

\subsubsection{Open boundary in $x$-direction}
With open boundary condition in $x$-direction, using properties of parabolic cylinder functions,
\begin{equation}
U(-\frac{\mu \varepsilon}{qB},    z ) \stackrel{ z \rightarrow \infty }{\rightarrow} 0,  \ \  \ \  U(-\frac{\mu \varepsilon}{qB},    -z ) \stackrel{ z \rightarrow \infty }{\rightarrow} \infty, \label{parabolicU}
\end{equation}
the coefficients $a=0$  and $d=0$. Also using Eq.(\ref{dGB1D}) and  relation
\begin{equation}
\mathcal{W} \{ U (-\frac{\mu \varepsilon}{qB},    z )  , U(-\frac{\mu \varepsilon}{qB},  -   z )  \} = \frac{\sqrt{2\pi}}{\Gamma(\frac{1}{2}-\frac{\mu \varepsilon}{qB})}, \label{wronskian}
\end{equation}
where $\mathcal{W} (f,g) = f g' - g f'$ stands for the Wronskian of two functions, so we obtain
\begin{align}
&G_B^{(Open,1D)}  (r_x  , r'_x  ; \varepsilon  ) = - 2\mu  \frac{\Gamma(\frac{1}{2}-\frac{\mu \varepsilon}{qB})}{  \sqrt{2qB } \sqrt{2\pi}}  \nonumber \\
& \times   U(-\frac{\mu \varepsilon}{qB},  - \sqrt{2qB } r_{x<})   U(-\frac{\mu \varepsilon}{qB},    \sqrt{2qB } r_{x>})  .
\end{align}

Considering another representation of open boundary condition 2D magnetic Green's function,
\begin{align}
& G_B^{(Open, k_y \mathbf{ e}_y ,2D)} (\bm{\rho},\bm{\rho}'; \varepsilon)   \nonumber \\
& =  \sum_{n_y } e^{- i k_y n_y L}    G_B^{( \infty, 2D)} (\bm{\rho} + n_y L \mathbf{ e}_y,\bm{\rho}'; \varepsilon)  , \label{G2Dopen}
\end{align}
we can also conclude that in addition to Eq.(\ref{G2Dinf}), another representation of $ G_B^{( \infty, 2D)}$ is given by
  \begin{align}
 &  G_B^{( \infty, 2D)} (\bm{\rho}  ,\bm{\rho}'; \varepsilon)  \nonumber \\
&  = - 2\mu  \frac{ \Gamma(\frac{1}{2}-\frac{\mu \varepsilon  }{qB}) }{  \sqrt{2qB } \sqrt{2\pi}}  \int_{-\infty}^\infty \frac{d p_y}{2\pi}  e^{i p_y  (r_y- r'_y  ) }        \nonumber \\
& \times  U(-\frac{\mu \varepsilon  }{qB},  - \sqrt{2qB } (r_{x <} + \frac{p_y}{qB} ))    \nonumber \\
& \times  U(-\frac{\mu \varepsilon   }{qB},    \sqrt{2qB } ( r_{x>} + \frac{p_y}{qB} ))   .
 \end{align}
Similar result and some interesting discussion of  quasi-classical approximation   of  $G_B^{( \infty, 2D)}$   can be found in \cite{pssb.2221300262}.

\subsubsection{Half open boundary in $x$-direction}
Next, let's consider only putting one hard wall on one side, {\it e.g.} 
 \begin{equation}
 G_B^{(Half, k_y \mathbf{ e}_y ,2D)} ( \bm{\rho}, \bm{\rho}' ;  \varepsilon)  |_{ r_{x>} \geqslant  \frac{n_q L}{2} \mathbf{ e}_x}  = 0 .
 \end{equation}
Using Eq.(\ref{parabolicU}) again, we can eliminating $U(-\frac{\mu \varepsilon}{qB},    \sqrt{2qB } r_{x<}) $ term in Eq.(\ref{G1Dgeneralsol}). The rest of coefficients can be determined by  implementing boundary condition and using Eq.(\ref{wronskian}) again, we thus find
\begin{align}
&G_B^{(Half, 1D)}  (r_x  , r'_x  ; \varepsilon  ) = - 2\mu  \frac{\Gamma(\frac{1}{2}-\frac{\mu \varepsilon}{qB})}{  \sqrt{2qB } \sqrt{2\pi}}     \nonumber \\
&  \times U(-\frac{\mu \varepsilon}{qB},  - \sqrt{2qB } r_{x<}) \bigg [  U(-\frac{\mu \varepsilon}{qB},    \sqrt{2qB } r_{x>})  \nonumber \\
&- \frac{U(-\frac{\mu \varepsilon}{qB},   \sqrt{2qB } \frac{L_+}{2}  )}{U(-\frac{\mu \varepsilon}{qB},  - \sqrt{2qB } \frac{L_+}{2}  )} U(-\frac{\mu \varepsilon}{qB},  - \sqrt{2qB } r_{x>}) \bigg ],
\end{align}
where
\begin{equation}
\frac{L_+}{2} = \frac{n_q L}{2} + \frac{p_y}{qB} .
\end{equation}

\subsubsection{Hard wall boundary in $x$-direction}
At last, let's consider   putting  hard walls on both sides, {\it e.g.} 
 \begin{equation}
 G_B^{(h.w., k_y \mathbf{ e}_y ,2D)} ( \bm{\rho}, \bm{\rho}' ;  \varepsilon)  |_{ r_{x<}  \leqslant  - \frac{n_q L}{2} \mathbf{ e}_x, \ r_{x>} \geqslant  \frac{n_q L}{2} \mathbf{ e}_x}  = 0 .
 \end{equation}
 Again,  implementing boundary condition and using Eq.(\ref{wronskian}), we   find
\begin{align}
&G_B^{(h.w., 1D)}  (r_x  , r'_x  ; \varepsilon  )  \nonumber \\
&= \frac{  2\mu  \frac{\Gamma(\frac{1}{2}-\frac{\mu \varepsilon}{qB})}{  \sqrt{2qB } \sqrt{2\pi}}   }{\frac{U(-\frac{\mu \varepsilon}{qB},  - \sqrt{2qB } \frac{L_+}{2}  ) }{U(-\frac{\mu \varepsilon}{qB},   \sqrt{2qB } \frac{L_+}{2}  )} -\frac{U(-\frac{\mu \varepsilon}{qB},   \sqrt{2qB } \frac{L_-}{2}  ) }{U(-\frac{\mu \varepsilon}{qB},  - \sqrt{2qB } \frac{L_-}{2}  )}  }   \nonumber \\
&  \times \bigg [ U(-\frac{\mu \varepsilon}{qB},  - \sqrt{2qB } r_{x<})   \nonumber \\
&  - \frac{U(-\frac{\mu \varepsilon}{qB},   \sqrt{2qB } \frac{L_-}{2}  ) }{U(-\frac{\mu \varepsilon}{qB},  - \sqrt{2qB } \frac{L_-}{2}  )}U(-\frac{\mu \varepsilon}{qB},  \sqrt{2qB } r_{x<})  \bigg ] \nonumber \\
& \times \bigg [  U(-\frac{\mu \varepsilon}{qB},    -\sqrt{2qB } r_{x>})  \nonumber \\
& - \frac{U(-\frac{\mu \varepsilon}{qB},  - \sqrt{2qB } \frac{L_+}{2}  ) }{U(-\frac{\mu \varepsilon}{qB},  \sqrt{2qB } \frac{L_+}{2}  )}U(-\frac{\mu \varepsilon}{qB},   \sqrt{2qB } r_{x>})  \bigg ],
\end{align}
where
\begin{equation}
\frac{L_-}{2} = \frac{n_q L}{2} - \frac{p_y}{qB} .
\end{equation}

\subsection{Spectrum of edge states with a contact interaction}
With a contact interaction, the quantization condition for various boundary conditions  are also given by the form of Eq.(\ref{qc2D}). In this section, we will simplify our discussion by setting $n_q=1$, hence only a single scatter is placed at origin. Even so, it is sufficient to demonstrate the difference between   edge state  solutions and bulk state solutions.

The magnetic zeta function for various boundary condition can be defined similarly to Eq.(\ref{zetafv}).  With only a single scatter placed at origin $(n_q=1)$, the generalized magnetic zeta functions for various boundary condition in $x$-direction  thus   all   have the form of
\begin{align}
&  \mathcal{M}^{( k_y \mathbf{ e}_y ,2D)}_B(\mathbf{ 0} , \mathbf{ 0}; \varepsilon)  \nonumber \\
 &= - \frac{4}{2\mu}  \frac{1}{L} \sum_{ p_y = \frac{2\pi  n_y}{L} +  k_y, n_y \in \mathbb{Z} }     G_B^{(1D)} (  \frac{p_y}{qB},   \frac{p_y}{qB} ; \varepsilon  ) \nonumber \\
& +   \frac{1}{\pi} (2 \gamma_E + \ln \frac{\mu \varepsilon \rho^2}{2}) |_{\rho \rightarrow 0} .
\end{align}

\subsubsection{Generalized magnetic zeta function for open boundary condition in $x$-direction}
The  generalized magnetic zeta functions  for open boundary condition in $x$-direction is thus  explicitly given by
\begin{align}
& \mathcal{M}^{(Open,  k_y \mathbf{ e}_y ,2D)}_B(\mathbf{ 0} , \mathbf{ 0}; \varepsilon)  \nonumber \\
&  = 4   \frac{\Gamma(\frac{1}{2}-\frac{\mu \varepsilon}{qB})}{  \sqrt{2qB } \sqrt{2\pi}}  \frac{1}{L} \sum_{ p_y = \frac{2\pi  n_y}{L} +  k_y, n_y \in \mathbb{Z} }   \nonumber \\
& \times   U(-\frac{\mu \varepsilon}{qB},  - \sqrt{2qB }  \frac{p_y}{qB})   U(-\frac{\mu \varepsilon}{qB},    \sqrt{2qB } \frac{p_y}{qB})   \nonumber \\
& +   \frac{1}{\pi} (2 \gamma_E + \ln \frac{\mu \varepsilon \rho^2}{2}) |_{\rho \rightarrow 0}. \label{zetaopen}
\end{align}
The infinite momentum sum in Eq.(\ref{zetaopen}) is UV divergent that is cancelled out by UV divergent part in the  second term. The   UV cancellation can be made explicitly by using Kummer  function representation of infinite volume magnetic Green's function and Eq.(\ref{zetaopen}), thus we find
 \begin{align}
& \mathcal{M}^{(Open,  k_y \mathbf{ e}_y ,2D)}_B(\mathbf{ 0} , \mathbf{ 0}; \varepsilon)   \nonumber \\
& =   \mathcal{M}^{( \infty ,2D)}_B(\mathbf{ 0} , \mathbf{ 0}; \varepsilon)  \nonumber \\
& +       \frac{ 1 }{ \pi}   \sum_{n_y \neq 0 } e^{- i k_y n_y L}  e^{- \frac{qB}{4}  |   n_y L \mathbf{ e}_y   |^2  }    \nonumber \\
& \times   \Gamma(\frac{1}{2} - \frac{\mu \varepsilon}{q B})   U (\frac{1}{2} - \frac{\mu \varepsilon}{q B}, 1,  \frac{qB}{2}  |   n_y L \mathbf{ e}_y |^2  )  , \label{Mzetaopen}
\end{align}
where $ \mathcal{M}^{( \infty ,2D)}_B$ is defined in Eq.(\ref{zetaMinf2D}).

\subsubsection{Generalized magnetic zeta function for half open boundary condition in $x$-direction}
For  half open boundary condition in $x$-direction, the UV divergence in infinite momentum sum can be regularized by subtracting by $ \mathcal{M}^{(Open,  k_y ,2D)}_B$, thus we find,
\begin{align}
&  \mathcal{M}^{(Half,  k_y \mathbf{ e}_y,2D)}_B(\mathbf{ 0} , \mathbf{ 0}; \varepsilon) \nonumber \\
&  =  \mathcal{M}^{(Open,  k_y\mathbf{ e}_y ,2D)}_B(\mathbf{ 0} , \mathbf{ 0}; \varepsilon)   - \frac{4}{2\mu}  \frac{1}{L} \sum_{ p_y = \frac{2\pi  n_y}{L} +  k_y, n_y \in \mathbb{Z} }  \nonumber \\
 & \times     \left [G_B^{(Half,1D)} (  \frac{p_y}{qB},   \frac{p_y}{qB} ; \varepsilon  ) - G_B^{(Open,1D)} (  \frac{p_y}{qB},   \frac{p_y}{qB} ; \varepsilon  )\right ] .
\end{align}

\subsubsection{Generalized magnetic zeta function for hard wall boundary condition in $x$-direction}
Similarly, for  hard wall boundary condition in $x$-direction,  by subtracting wtih $ \mathcal{M}^{(Open,  k_y ,2D)}_B$,   the UV regularized magnetic zeta function is given by
\begin{align}
&  \mathcal{M}^{(h.w.,  k_y \mathbf{ e}_y ,2D)}_B(\mathbf{ 0} , \mathbf{ 0}; \varepsilon) \nonumber \\
&  =  \mathcal{M}^{(Open,  k_y \mathbf{ e}_y,2D)}_B(\mathbf{ 0} , \mathbf{ 0}; \varepsilon)   - \frac{4}{2\mu}  \frac{1}{L} \sum_{ p_y = \frac{2\pi  n_y}{L} +  k_y, n_y \in \mathbb{Z} }  \nonumber \\
 & \times     \left [G_B^{(h.w.,1D)} (  \frac{p_y}{qB},   \frac{p_y}{qB} ; \varepsilon  ) - G_B^{(Open,1D)} (  \frac{p_y}{qB},   \frac{p_y}{qB} ; \varepsilon  )\right ] . \label{Mzetahardwall}
\end{align}

  \begin{figure}
\begin{center}
\includegraphics[width=1.0\textwidth]{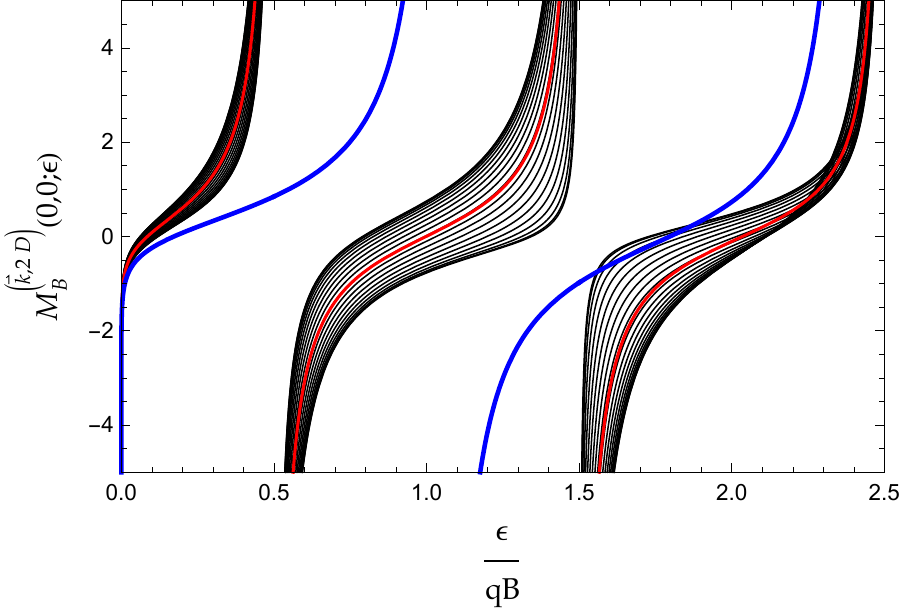}
\caption{   Bulk energy bands (filled with black curves) in  finite volume with magnetic periodic boundary condition in both $x$- and $y$-direction is generated by    varying $k_x$ in magnetic Brillouin zone:  $k_x \in [0 , \frac{2\pi}{n_q L}]$ in finite volume magnetic zeta function $ \mathcal{M}^{( L, \mathbf{ k} ,2D)}_B(\mathbf{ 0} , \mathbf{ 0}; \varepsilon) $  defined in Eq.(\ref{zetafv}). The wave vector in $y$-direction,  $k_y$, is fixed at $k_y= \frac{\pi}{2 n_q L}$.  Compared with    $ \mathcal{M}^{(Open,  k_y \mathbf{ e}_y ,2D)}_B(\mathbf{ 0} , \mathbf{ 0}; \varepsilon)  $ in Eq.(\ref{Mzetaopen}) (red curve) with open boundary condition along $x$-direction  and   $\mathcal{M}^{(h.w.,  k_y \mathbf{ e}_y ,2D)}_B(\mathbf{ 0} , \mathbf{ 0}; \varepsilon)$ in Eq.(\ref{Mzetahardwall}) (blue curve) with a hard wall boundary condition in $x$-direction. The parameters are chosen as:      $L=5$, and $n_q=n_p=1$.}\label{Mzetaplot}
\end{center}
\end{figure}

\subsubsection{Edge states vs. bulk energy bands}
In general, the energy spectrum for various boundary conditions must be generated by using   Eq.(\ref{qc2D}). The topological edge states in gaps between allowed energy bands in fact can be illustrated by only considering a  simple case with  $n_q=1$. For a single contact interaction in the box with $n_q=1$,   the quantization condition is thus simply given by
\begin{equation}
\cot \delta_0 (\varepsilon) =  \mathcal{M}^{(  2D)}_B(\mathbf{ 0} , \mathbf{ 0}; \varepsilon) .
\end{equation}
For periodic boundary conditions in both $x$-and $y$-direction, with a fixed $k_y$, the bulk energy bands can be generated by treating $k_x$   as a free parameter in finite volume magnetic zeta function $ \mathcal{M}^{(L,  \mathbf{ k} ,2D)}_B(\mathbf{ 0} , \mathbf{ 0}; \varepsilon) $  that is defined in Eq.(\ref{zetafv}), see Fig.~\ref{Mzetaplot}. The bulk energy bands are separated by gaps in betweens. The edge states are produced by replacing the periodic boundary condition in $x$-direction  by a hard wall boundary condition,   $\mathcal{M}^{(h.w.,  k_y \mathbf{ e}_y ,2D)}_B(\mathbf{ 0} , \mathbf{ 0}; \varepsilon)$ in Eq.(\ref{Mzetahardwall}).  As shown in Fig.~\ref{Mzetaplot}, unlike topologically trivial edge states,   the solutions of edge states of a magnetic system not only show up in the gap, but also punch through bulk energy bands.

\section{Analytic properties of finite volume solutions}\label{analyticproperty}

The periodicity of lattice structure and particles interaction create the band structures, the energy spectrum split into isolated bands separated by gaps  in betweens. Hence, when wave vector $\mathbf{ k}$ is changed continuously, the energy of particle must experience a discontinuity when particle jumps from one band to another. It has been shown \cite{PhysRev.115.809,Heine_1963}  that when wave vector is taken complex at the edge of Brillouin zone, the real energy solutions in the gap are   possible, hence the transition from one band to another can be made smoothly in complex $\mathbf{ k}$ plane. The complex wave vector  at the edge of Brillouin zone may be interpreted as edge solutions with a penetrable wall on the edge or surface of material. In  presence of a magnetic field, the real energy solutions can also be found for complex wave vector, however, the situation is more complicated, the energy solutions not only appear in the gap but also penetrate into bulk energy bands because of non-trivial topology.

  \begin{figure}
\begin{center}
\includegraphics[width=1.0\textwidth]{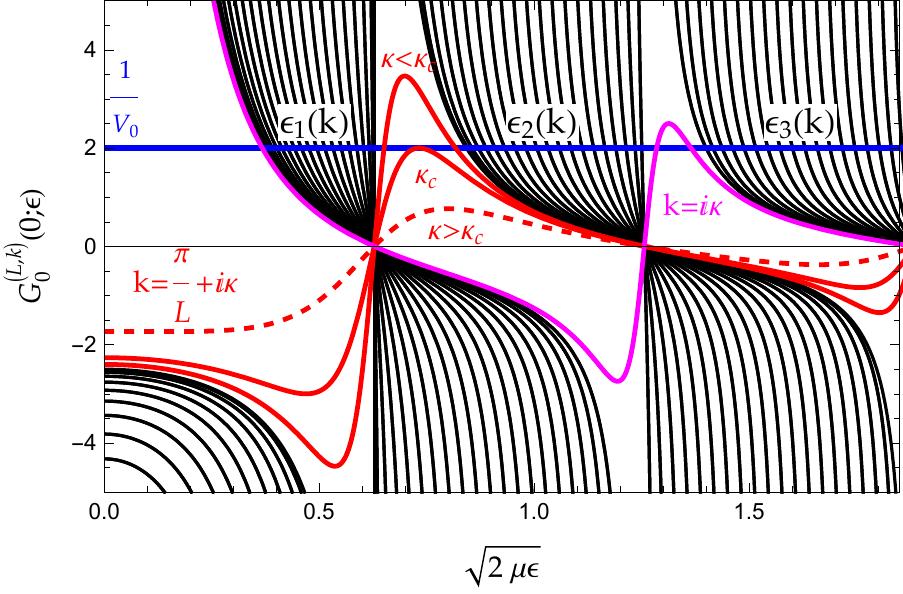}
\caption{   Plot of  $G^{(L,k)}_0 (0 ;\varepsilon)$  defined in Eq.(\ref{G1Dfv}) vs. $\frac{1}{V_0}$ (blue solid line):  the area of allowed energy bands are filled by  black curves (real $k \in [0, \frac{2\pi}{L}]$ values), the energy bands are separated by gaps; The red and purple curves  that show up in   gaps are generated   with complex wave vectors $k= \frac{\pi}{L} + i \kappa$ and $k=  i \kappa$ respectively. A pair of energy solutions in the gap can be found for $\kappa < \kappa_c$. The parameters are: $V_0=0.5$, $\mu=1$  and $L=5$.}\label{GLV0plot}
\end{center}
\end{figure}

  \begin{figure}
\begin{center}
\includegraphics[width=1.0\textwidth]{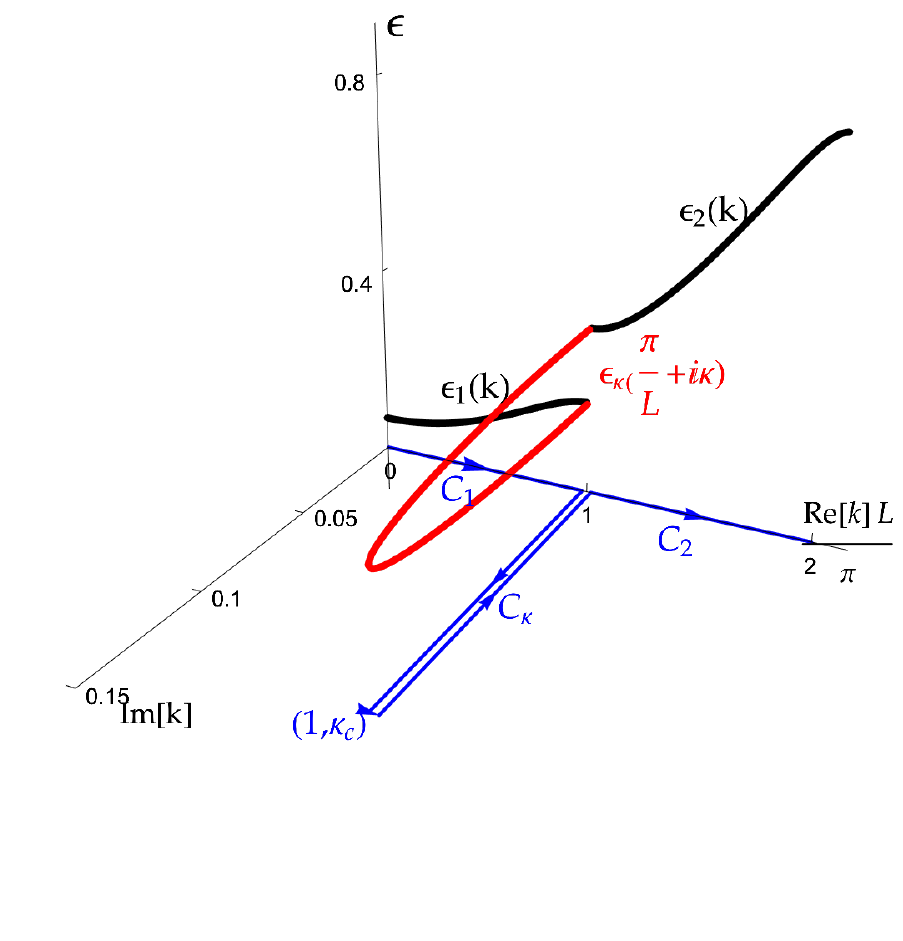}
\caption{   The motion of $\varepsilon(k)$ as the wave vector $k$ moves continuously in complex plane follow the path: $C_1 \rightarrow C_\kappa \rightarrow C_2$.  $C_1$ and $C_2$ are on real axis in $k$ plane between $[0, \frac{\pi}{L}]$ and $[\frac{\pi}{L}, \frac{2\pi}{L}]$ respectively.  $C_\kappa$ is on complex plane with $k = \frac{\pi}{L} + i \kappa$, $\kappa: 0 \rightarrow \kappa_c \rightarrow 0$. The energy solutions thus moves continuously from $\varepsilon_1 (k)$ into $\varepsilon_\kappa (\frac{\pi}{L} + i \kappa)$ in the gap, and then connected into $\varepsilon_2 (k)$.  }\label{bandcontourplot}
\end{center}
\end{figure}

In this subsection, we first give a brief summary of complex wave vector with a simple 1D example. With a contact interaction, the quantization condition in 1D is given by
\begin{equation}
\frac{1}{V_0} = G^{(L,k)}_0 (0; \varepsilon),
\end{equation} 
where 1D finite volume Green's function is   given by
\begin{align}
\frac{1}{2\mu}G^{(L,k)}_0 (0 ;\varepsilon) &= -  \sum_{n \in \mathbb{Z}} \frac{ i e^{i \sqrt{ 2\mu \varepsilon} | n L| }  }{\sqrt{2\mu \varepsilon}} e^{- i k n L}    \nonumber \\
&= -  \frac{  2\mu}{2 \sqrt{2\mu \varepsilon}} \frac{\sin \sqrt{2\mu \varepsilon} L }{\cos \sqrt{2\mu \varepsilon} L -   \cos  k L  }  . \label{G1Dfv}
\end{align}
The finite volume Green's function $G^{(L,k)}_0  $ remains real as for the real value of $k$, which yields the real dispersion relation of 
\begin{equation}
\varepsilon=\varepsilon (k)=\varepsilon (-k) = \varepsilon (k + \frac{2\pi}{L}).
\end{equation}
 The band structures is explicitly  produced by the bound of $| \cos  k L | \leqslant 1$.  Using Eq.(\ref{G1Dfv}), one can show that for $k= \frac{\pi d}{L} + i \kappa$, $d \in \mathbb{Z}$,  Green's function is still a real function,
  \begin{equation}
 G^{(L, \frac{\pi d}{L} + i \kappa )}_0 (0 ;\varepsilon)  = -  \frac{  2\mu}{2 \sqrt{2\mu \varepsilon}} \frac{\sin \sqrt{2\mu \varepsilon} L }{\cos \sqrt{2\mu \varepsilon} L - (-1)^d \cosh  \kappa L  } .
\end{equation}
Hence, we can see clearly because of $$ \cosh  \kappa L \geqslant 1 ,$$   the energy  solutions of complex wave function, $k= \frac{\pi d}{L} + i \kappa$, only show up in the gaps between bands, see Fig.~\ref{GLV0plot}. In the gaps, for a fixed $V_0$, a pair of energy solutions can be found for finite value of $\kappa$. The gap between two solutions shrinks when $\kappa $ is increased, until $\kappa$ reach its critical point $\kappa_c$, the gap close up, two solutions becomes degenerate. Beyond $\kappa_c$, no solutions can be found, see Fig.~\ref{GLV0plot} as an example. 
Therefore, the complex wave vector can be used as a parameter to navigate across bulk energy bands smoothly. Using  Fig.~\ref{bandcontourplot} as an example, two allowed energy bands $\varepsilon_1 (k)$ and  $\varepsilon_2 (k)$  are separated by a gap for real values of $k$'s. Imaging  wave vector $k$ start at $k=0$, and is forced to move following the path of $C_1 \rightarrow C_\kappa \rightarrow C_2$ in  Fig.~\ref{bandcontourplot},   where  both $C_1$ and $C_2$ are defined on real axis for $k \in [0, \frac{\pi}{L}]$ and  $k \in [ \frac{\pi}{L}, \frac{2\pi}{L}]$  respectively.  The contour $C_\kappa$ is defined in complex $k$ plane with fixed   $Re[k]=\frac{\pi}{L}$ value,  and the imaginary part of $Im[k] = \kappa$ is circling around $\kappa_c$, $\kappa: 0 \rightarrow \kappa_c \rightarrow 0$. While $k$ is on $C_1$, the energy solution stays in energy band $\varepsilon_1 (k)$  following the motion of $k$, moving from lower edge $\varepsilon_1 (0)$ up to upper edge  $\varepsilon_1 (\frac{\pi}{L})$. While $k$ is extended into complex plane on $C_\kappa$, the energy solution then protrude into the gap between two allowed bands, and continue climbing up to the lower edge of energy band $\varepsilon_2 (k)$ at $\varepsilon_2 (\frac{\pi}{L})$. Then, it merged into second band $\varepsilon_2 (k)$  if $k$ is increased further on  $C_2$. Similarly,  $\varepsilon_2 (k)$ and $\varepsilon_3 (k)$ are smoothly connected by taking wave vector into complex plane at the edge of Brillouin zone: $k= \frac{2 \pi }{L} + i \kappa$ which is equivalent to $k=  i \kappa$, see Fig.~\ref{GLV0plot}. We can also see from Fig.~\ref{GLV0plot} that the curves of $ G^{(L, \frac{\pi d}{L} + i \kappa )}_0 (0 ;\varepsilon) $ with different $d$ values occupy completely different territories, hence there are no denegerate energy solutions for different  $d$ values.

  \begin{figure}
\begin{center}
\includegraphics[width=1.0\textwidth]{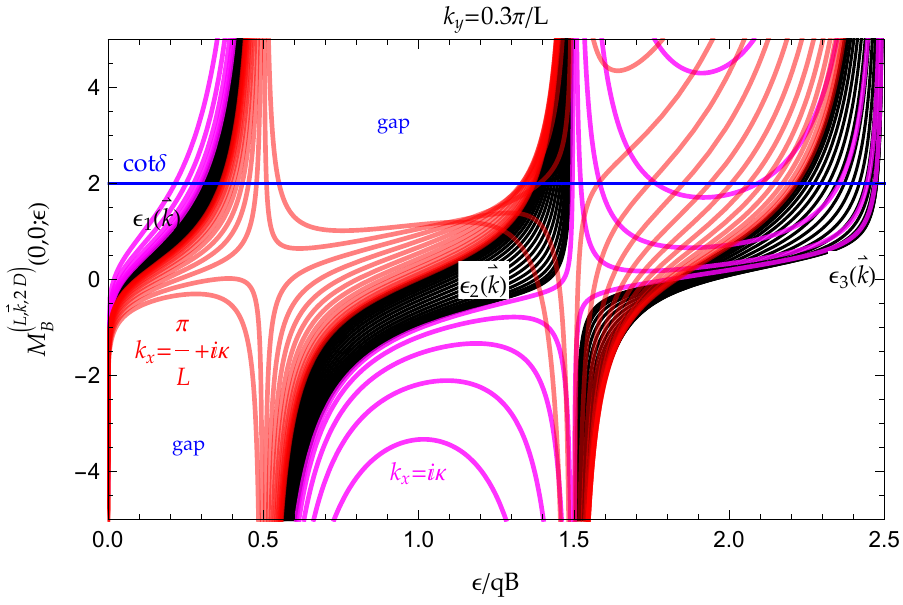}
\includegraphics[width=1.0\textwidth]{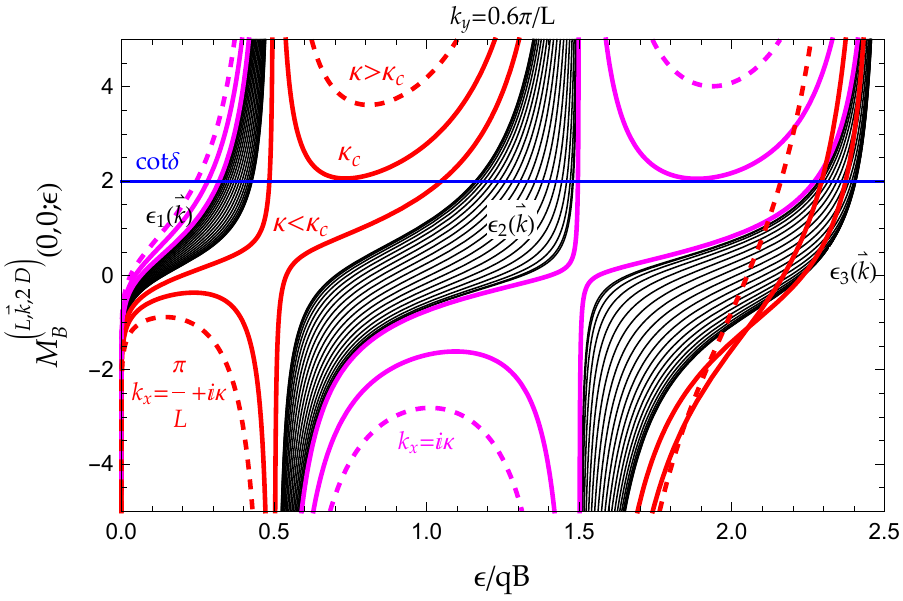}
\caption{   Energy band structure  (filled with black curves)  is the result of periodicity in both $x$- and $y$-direction, red $(d=1)$ and purple ($d=0$) curves are generated by taking $k_x$ into complex plane: $k_x = \frac{\pi d}{L} + i \kappa$.  $k_y$'s are fixed at $k_y= 0.3 \frac{\pi}{   L}$ and $k_y= 0.6 \frac{\pi}{   L}$ in upper and lower panels respectively. The parameters are chosen as:   $L=5$, and $n_q=n_p=1$. Blue line represents a constant $\cot \delta_0 (\varepsilon)$ that is used  only to help to visualize the energy solutions.}\label{edgeplot}
\end{center}
\end{figure}

  \begin{figure}
\begin{center}
\includegraphics[width=1.0\textwidth]{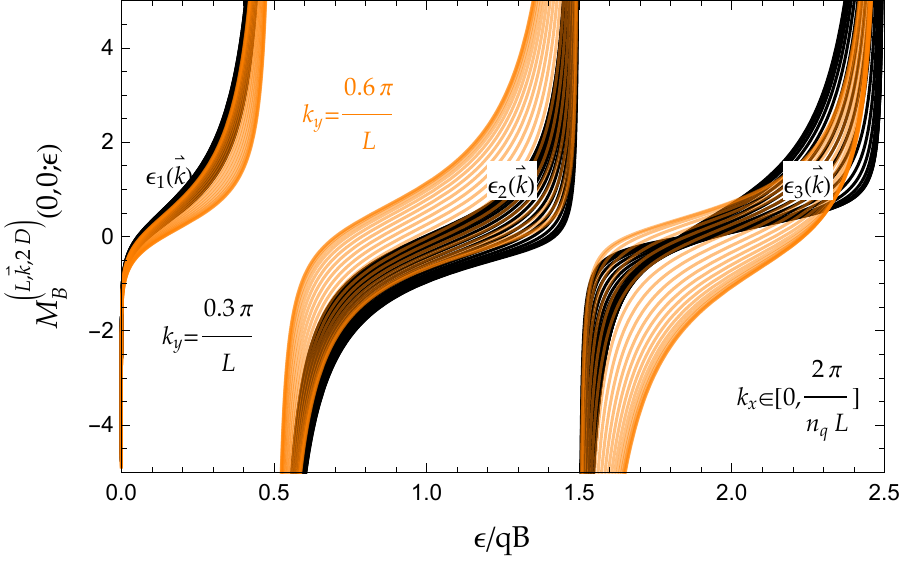}
\caption{   Overlapping energy bands  of different $k_y$ values: $k_y=0.3 \frac{\pi}{L}$ (black) vs. $k_y=0.6 \frac{\pi}{L}$ (orange) .}\label{twobandsplot}
\end{center}
\end{figure}

In presence of magnetic field,  with a complex wave vector
\begin{align}
\mathbf{ k} = (\frac{\pi d}{n_q L} + i \kappa) \mathbf{ e}_x + k_y \mathbf{ e}_y, \ \ \ \ d \in \mathbb{Z},
\end{align}
similarly, the real energy solutions are also available, however situation becomes much more intriguing. Unfortunately, for a magnetic system, analytic properties cannot be shown easily in a straightforward way, all the discussions heavily rely on numerics.  Let's consider the case of $n_q=1$ as a simple example,  which corresponds to a single contact interaction placed at origin, thus the magnetic zeta function is given by
\begin{align}
& \mathcal{M}^{(L,(\frac{\pi d}{n_q L} + i \kappa) \mathbf{ e}_x + k_y \mathbf{ e}_y ,2D)}_B( \mathbf{ 0} ,  \mathbf{ 0}; \varepsilon)    \nonumber \\
&=   \mathcal{M}^{( \infty, 2D)}_B( \mathbf{ 0} ,  \mathbf{ 0};  \varepsilon)   \nonumber \\
& +  \frac{ 1 }{\pi}     \sum_{\mathbf{ n}_B \neq \mathbf{ 0}} (-1)^{d n_x } (-1)^{n_p n_x n_y}   e^{ \kappa n_x n_q L } e^{- i k_y n_y L}     \nonumber \\
&     \times  e^{- \frac{qB}{4}  |  \mathbf{ n}_B L   |^2  }   \Gamma(\frac{1}{2} - \frac{\mu \varepsilon}{q B})  U (\frac{1}{2} - \frac{\mu \varepsilon}{q B}, 1,  \frac{qB}{2}  |   \mathbf{ n}_B L   |^2  )   ,
\end{align}
which is indeed a real function. Compared with previously discussed 1D topologically trivial example, there are some new features in a magnetic system.  First of all, as we can see in Fig.~\ref{edgeplot}, for small   $k_y$, the gap area between allowed energy bands  cannot be completely filled by taking $k_x$ into complex plane, see gap between $\varepsilon_1$ and $\varepsilon_2$ bands in upper panel in Fig.~\ref{edgeplot}. Hence, for certain range of  $k_y$,  although complex wave vector $k_x$ may narrow the gap, the gap   remains. Therefore, using complex  $k_x$ alone to navigate though gaps are not possible for certain range of $k_y$, however, due to overlapping energy bands of different $k_y$, see Fig.~\ref{twobandsplot}, it may be still possible by using both complex $k_x$ and real $k_y$ to navigate through different energy bands smoothly by avoiding gap area. Secondly, with complex wave vector $k_x=\frac{\pi d}{n_q L} + i \kappa$,   curves  not only show up  in the gap areas, some curves   punch through the allowed bulk bands, and invade into the gap areas with different $d$ value, see Fig.~\ref{edgeplot}. In addition, the curves with complex wave vectors in gap make up a vortex shape, all the curves are pushed away from a vortex centered at location of Landau level energy: $\varepsilon_n=\frac{qB}{\mu} (n + \frac{1}{2})$, see example  in Fig.~\ref{edgeplot}. These irregular behaviors of magnetic zeta function with a complex wave vector may have a topological origin.

\section{Summary}\label{summary}

 In summary,  we explore and discuss  some topological and analytic properties of a finite volume two-body system in a uniform magnetic field   in present work.    The Berry phase is introduced on a torus of magnetic Brillouin zone in $\mathbf{ k}$-space. The analytic solutions of edge states with a hard wall boundary condition in $x$-direction are also presented and discussed.  By further taking  $\mathbf{ k}$ into a complex plane, the analytic properties of energy spectrum is also discussed.

\acknowledgments

We   acknowledges support from the Department of Physics and Engineering, California State University, Bakersfield, CA.

\appendix

\bibliography{ALL-REF.bib}

\end{document}